\documentclass[twocolumn, dvipsnames]{aastex631}
\usepackage[autostyle]{csquotes}
\MakeOuterQuote{"}
\usepackage{hyperref}
\hypersetup{
    colorlinks=true,
    linkcolor=Thistle,
    filecolor=Thistle,      
    urlcolor=Thistle,
    citecolor=Thistle,
    }
\usepackage[T1]{fontenc}
\usepackage{graphicx}	% Including figure files
\usepackage{amsmath}	% Advanced maths commands

\newcommand{\gaia}{{Gaia}}

\newcommand{\kepler}{{Kepler}}
\newcommand{\ktwo}{K2}
\newcommand{\dnu}{$\Delta\nu$}
\newcommand{\numax}{$\nu_{\rm max}$}
\newcommand{\ktwogap}{\ktwo\ GAP}
\newcommand{\feh}{${\rm [Fe/H]}$}

\defcitealias{warfield+2021}{War21}
\defcitealias{APOKASC2}{Pin18}
\defcitealias{catalogpaper}{JSS24}
\newcommand{\akccite}{\citetalias{catalogpaper}}

\received{December 21, 2023}
\revised{March 9, 2021}
\accepted{March 10, 2024}

\shorttitle{The APO-K2 Catalog II}
\shortauthors{Warfield et al.}
\graphicspath{{./}{figures/}}
\begin{document}

\title{The APO-K2 Catalog. II. Accurate Stellar Ages for Red Giant Branch Stars Across the Milky Way}

\correspondingauthor{Jack T. Warfield}
\email{jtw5zc@virginia.edu}

\author[0000-0003-1634-4644]{Jack T. Warfield}
\affiliation{Department of Astronomy, The University of Virginia, 530 McCormick Road, Charlottesville, VA 22904, USA}

\author[0000-0002-7550-7151]{Joel C. Zinn}
\affiliation{Department of Physics and Astronomy, California State University Long Beach, Long Beach, CA 90840, USA}

\author[0000-0002-1043-8853]{Jessica Schonhut-Stasik}
\altaffiliation{Neurodiversity Inspired Science and Engineering Graduate Fellow}
\affiliation{Department of Physics \& Astronomy, Vanderbilt University 6301 Stevenson Center Ln, Nashville, TN 37235, USA}

\author[0000-0002-6534-8783]{James W. Johnson}
\affiliation{The Observatories of the Carnegie Institution for Science, 813 Santa Barbara St., Pasadena, CA, 91101, USA}
\affiliation{Department of Astronomy, The Ohio State University, 140 W 18th Ave, Columbus, OH 43210, USA}
\affiliation{Center for Cosmology and AstroParticle Physics, The Ohio State University, 191 W Woodruff Ave, Columbus, OH 43210, USA}

\author[0000-0002-7549-7766]{Marc H. Pinsonneault}
\affiliation{Department of Astronomy, The Ohio State University, 140 W 18th Ave, Columbus, OH 43210, USA}
\affiliation{Center for Cosmology and AstroParticle Physics, The Ohio State University, 191 W Woodruff Ave, Columbus, OH 43210, USA}

\author[0000-0001-7258-1834]{Jennifer A. Johnson}
\affiliation{Department of Astronomy, The Ohio State University, 140 W 18th Ave, Columbus, OH 43210, USA}
\affiliation{Center for Cosmology and AstroParticle Physics, The Ohio State University, 191 W Woodruff Ave, Columbus, OH 43210, USA}

\author[0000-0002-4879-3519]{Dennis Stello}
\affiliation{School of Physics, University of New South Wales, NSW 2052, Australia}
\affiliation{Sydney Institute for Astronomy (SIfA), School of Physics, University of Sydney, NSW 2006, Australia}
\affiliation{Stellar Astrophysics Centre, Department of Physics and Astronomy, Aarhus University, DK-8000 Aarhus C, Denmark}

\author[0000-0002-1691-8217]{Rachael L. Beaton}
\affiliation{Space Telescope Science Institute, Baltimore, MD 21218, USA}
\affiliation{Department of Physics and Astronomy, Johns Hopkins University, Baltimore, MD 21218, USA}

\author{Yvonne Elsworth}
\affiliation{Stellar Astrophysics Centre, Department of Physics and Astronomy, Aarhus University, Ny Munkegade 120, DK-8000 Aarhus C, Denmark}
\affiliation{School of Physics and Astronomy, University of Birmingham, Edgbaston, Birmingham, B15 2TT, UK}

\author[0000-0002-8854-3776]{Rafael A. Garc\'ia}
\affiliation{Universit\'e Paris-Saclay, Universit\'e Paris Cit\'e, CEA, CNRS, AIM, 91191, Gif-sur-Yvette, France}

\author[0000-0002-0129-0316]{Savita Mathur}
\affiliation{Instituto de Astrof\'isica de Canarias (IAC), E-38205 La Laguna, Tenerife, Spain}
\affiliation{Universidad de La Laguna (ULL), Departamento de Astrof\'isica, E-38206 La Laguna, Tenerife, Spain}

\author[0000-0002-7547-1208]{Beno\^it Mosser}
\affiliation{LESIA, Observatoire de Paris, Universit\'e PSL, CNRS, Sorbonne Universit\'e, Universit\'e de Paris, 92195 Meudon, France}

\author[0000-0001-6359-2769]{Aldo Serenelli}
\affiliation{Institute of Space Sciences (ICE, CSIC), Carrer de Can Magrans S/N, Campus UAB, E-08193 Bellaterra, Spain}
\affiliation{Institut d'Estudis Espacials de Catalunya, Carrer Gran Capitá 2, E-08034, Barcelona, Spain}

\author[0000-0002-4818-7885]{Jamie Tayar}
\affiliation{Department of Astronomy, University of Florida, Bryant Space Science Center, Stadium Road, Gainesville, FL 32611, USA}



\begin{abstract}

We present stellar age determinations for 4,661 red giant branch (RGB) stars in the APO-K2 Catalog, derived using mass estimates from K2 asteroseismology from the K2 Galactic Archaeology Program and elemental abundances from the Apache Point Galactic Evolution Experiment (APOGEE) survey. Our sample includes 17 of the 19 fields observed by K2, making it one of the most comprehensive catalogs of accurate stellar ages across the Galaxy in terms of the wide range of populations spanned by its stars, enabling rigorous tests of Galactic chemical evolution models. Taking into account the selection functions of the K2 sample, the data appear to support the age-chemistry morphology of stellar populations predicted by both inside-out and late-burst scenarios. We also investigate trends in age versus stellar chemistry and Galactic position, {which are consistent with previous findings}. Comparisons against APOKASC-3 asteroseismic ages show agreement to within $\sim$3\%. We also discuss offsets between our ages and spectroscopic ages. Finally, we note that ignoring the effects of $\alpha$-enhancement on stellar opacity {(either directly or with the Salaris metallicity correction)} results in an $\sim$10\% offset in age estimates for the most $\alpha$-enhanced stars, which is an important consideration for continued tests of Galactic models with this and other asteroseismic age samples.

\end{abstract}

\keywords{Stellar ages (1581) --- Asteroseismology (73) --- Stellar abundances (1577) --- Milky Way evolution (1052) --- Milky Way formation (1053) --- Galaxy stellar content (621) --- Red giant stars (1372) --- Catalogs (205) --- Stellar evolutionary models (2046)}


%Styles:
%\begin{itemize}
%    \item American English: disk, catalog, gray
%    \item AAS style, so no telescope italics (e.g., \kepler)
%    \item $\alpha$ bi-modality
%    \item K2 GAP
%    \item APOKASC-2, APOKASC-3
%    \item tt or texttt, not smallcaps or sc for package names, etc.
%\end{itemize}

\section{Introduction} \label{sec:intro}

% The importance of deriving stellar ages for Galactic evolution. 
%Even today, we are able to observe and study some of the oldest stars to have formed in our Galaxy, and we already have clear evidence for trends between the ages, compositions, motions, and positions of stars. Studied more closely, therefore, and in greater detail, being able to determine the ages of stars in our Galaxy may be vital for attaining a deep understanding of the the Milky Way, both past and present.
%Deriving the ages of stars in our Galaxy is one of the most vital calculation for our continued understanding of stellar and Galactic evolution. Ages are particularly useful for red giant stars, as they are luminous at large distances, allowing for age distributions to be determined throughout the Galaxy. Furthermore, sun-like oscillations, necessary for the derivation of ages through asteroseismology are easier to obtain for red giant stars because they can be observed in a longer cadence than similar oscillations in main sequence stars. 
% Why it is hard to derive stellar ages. 
%However, calculating stellar ages, as they are not a directly measurable quantity, has proved difficult. Often, studies rely on abundance data such as low absolute iron abundance or characteristic heavy element abundance patterns relative to iron to indicate the ages of stellar populations. 
The complex formation history of the Milky Way (MW) is both important to understand and difficult to decode. Stellar ages are a crucial clue, complementing studies of stellar positions, dynamics, and composition.
However, due to the fact that the observed properties of stars are mostly insensitive to age, precise and accurate ages of stars are difficult to infer.
%% Why asteroseismology is important
%Because ages are not a directly measurable quantity, it is difficult to precisely and accurately calculate the age of an individual star \citep{soderblom2010}. 
% Marc: This really depends - I think it's more accurate to say that ages are challenging to infer for many stars because their observed properties are insensitive to age. Asteroseismology is incredibly useful because it adds an observable - mass - that is a potent chronometer.

With the rise of ensemble asteroseismology over the last two decades---thanks to large-scale, space-based, time-domain surveys such as CoRoT \citep{corotoverview}, Kepler \citep{borucki+2010}, K2 \citep{howell+2014}, and TESS \citep{Ricker2015}---it is possible to measure solar-like oscillation patterns in many stars.
%it is possible to calculate precise ages for large numbers of stars using measurements of their intrinsic oscillations. % Marc: it is possible to measure stellar oscillation patterns in many stars,
These oscillations are due to near-surface turbulent convection motions that generate sound waves.
% Savita: This is the case of solar-like oscillations in stars with a radiative interior and an outer convective zone.
These sound waves, when at distinct resonant frequencies in the stellar interior, create standing waves that form a frequency pattern of overtone modes of differing spherical degree and radial order. The characteristic spacing between these frequencies, known as the large frequency-spacing ($\Delta\nu$), is related to the mean density of the star \citep{tassoul1980, kjeldsenbedding1995}. The frequency of maximum acoustic power ($\nu_{\rm max}$) is related to the acoustic cutoff frequency, and therefore the density scale height and surface gravity of the star \citep{brown1991, kjeldsenbedding1995}. These global parameters allow us to derive the masses of the stars through well-understood scaling relations as long as these two parameters and the stars' effective surface temperatures ($T_{\mathrm{eff}}$) are known. These mass measurements, along with composition information, allow model-based age determinations.

% The importance of spectroscopy
%Asteroseismology can provide incredible precise measurements of $\Delta\nu$ and $\nu_{\mathrm{max}}$, leaving T$_{\mathrm{eff}}$ as the limiting parameter for age measurements. However, the increasing precision of spectroscopic temperatures allows us to improve our age estimations for a wealth of stars in the Galaxy. 

% Missions
%Asteroseismic data have been made abundant thanks to large, space-based, time-domain surveys such as \textit{COROT} \citep{corotoverview}, \textit{Kepler} \citep{borucki+2010}, \ktwo\ \citep{howell+2014}, and {\it TESS} \citep{Ricker2015}. The asteroseismic targets from these misssions have been further supported by, with significant cross-matches, similarly large spectroscopic surveys such as APOGEE \citep{majewski2017}, LAMOST \citep{Cui2012}, and GALAH \cite{DeSilva2015}. The first example of combining these large data sets of asteroseismology and spectroscopy is the work of \citet{apokasc1, APOKASC2} which produced catalogs of asteroseismic global parameters, asteroseismic masses and radii, and age estimates in the \textit{Kepler} field by combining \textit{Kepler} asteroseismic measurements with spectroscopy from APOGEE. 
Analyses from asteroseismic data have been significantly furthered due to support from similarly large, ground-based, spectroscopic surveys such as the Apache Point Galactic Evolution Experiment (APOGEE; \citealt{majewski2017}), The Large Sky Area Multi-Object Fibre Spectroscopic Telescope (LAMOST) survey \citep{Cui2012}, and the GALactic Archaeology with HERMES (GALAH) survey \citep{DeSilva2015}. These surveys, in select cases, have intentionally large overlaps with targets from space-based asteroseismic missions. The resulting ages due to the combination of spectroscopic compositions and temperatures with asteroseismic masses have allowed significant work in Galactic archaeology, revealing the Galaxy's evolution history by linking stellar chemistry and age at Galaxy-wide scales (e.g., \citealt{Anders2017}, \citealt{SilvaAguirre+2018}, {\citealt{rendle+2019},} \citealt{Miglio2021}, \citealt{willett+2023}, \citealt{imig+2023}, \citealt{stokholm+2023}).

% Alpha Elements
%Our understanding of the formation of the Milky Way can be examined using the existence of $\alpha$-rich stars \citep{aller1960, wallerstein1962}. These stars are rich in elements such as O, Mg, Si, S, Ca, and Ti; elements primarily produced in core-collapse supernova (SNe II). The nature of these supernova are short-lived, indicating that when these elements are seen in abundance these stars must be old. Conversely, Fe-peak elements are created in both SNeII and Type Ia super --- which require a longer-lived progenitor, and their resulting stars to have more Fe-peak abundances. 
% Marc: Need more background.  Suggest "The mix of heavy elements in stars is not universal; instead, it arises from distinct sources with different timescales. As a result, regions with rapid star formation will have a different mix than ones from regions with more gradual, or episodic, star formation."
One avenue through which the formation of the MW can be examined with stellar compositions is by comparing populations of $\alpha$-rich versus $\alpha$-poor stars (as discussed by, e.g., \citealt{aller1960} and \citealt{wallerstein1962}). These stars are rich (or poor) in $\alpha$-capture elements (e.g., Mg, O) compared to the Sun. {The mix of heavy elements in stars is not universal; instead, it arises from distinct sources contributing on different timescales. As a result, regions with rapid star formation will have a different mix of elements versus regions with more gradual, or episodic, star formation.} $\alpha$ elements are primarily produced in core-collapse supernova (SNe II), which---owing to their massive, short-lived progenitors---have rates that closely track the Galaxy's (at least the local) star formation history (SFH). SNe Ia, a significant source of iron-peak elements, only enrich the interstellar medium at later times due a combination of the longer lifetime of intermediate-mass SNe Ia progenitors and delay time distributions from Chandrasekhar mass overflow \citep{timmes_woosley_weaver1995,kobayashi+1998,ruiter_belczynski_fryer2009}.
% How this turns up in models
%When modeling stars in our Galaxy, and considering the different types of supernova, as the Galaxy evolves the ratio of [$\alpha$/Fe] would decrease as the frequency of SNe Ia increases, as the Fe-peak elements become more abundant. Eventually these would reach an equilibrium ration (e.g., \citep{Weinberg+2017}). However, this expected trend is not observed in the solar neighbourhood \citep{prochaska2000, Bensby2003}; rather stars with -1 $<$ [Fe/H] $<$ 0 present with a range of $\alpha$ values, that has become known as the $alpha$-bi-modality, due to its bi-modal distribution in the [$\alpha$/H] vs. [Fe/H] space. Investigations have led to further discoveries related to the $\alpha$-bi-modality, such as distinct trends in [$\alpha$/Fe] vs. [Fe/H] for the geometrically defined thin and thick disks \citep{Bensby2003, GilmoreReid1983}. \citet{Hayden2015} found that the high-$\alpha$ part of this sequence presents most significantly at a particular height above the Galactic plane (|Z| $>$ 0.5 kpc) and a certain Galactic radius (R $<$ 11 kpc), using $~$7,000 red giants from the Sloan Digital Sky Survey (SDSS) Apache Point Galactic Evolution Experience (APOGEE) Data Release 12 (DR12, \citep{Alam2015}). 
In models of Galactic star formation, as the Galaxy evolves and the rate of SNe Ia increases, the predicted ${\rm [\alpha/Fe]}$\footnote{${\rm [\alpha/Fe]} = \log_{10}{(N_{\alpha}/N_{\rm Fe})} - \log_{10}{(N_{\alpha}/N_{\rm Fe})_\odot}$.} of new stars simultaneously decreases as the Fe-peak elements become more abundant. Eventually, an equilibrium ratio is reached (e.g., \citealt{Weinberg+2017}).

Regardless, this expected trend is not observed in the solar neighbourhood \citep{prochaska2000, Bensby2003}; rather, stars with $-1 < {\rm [Fe/H]} < 0$ present a discontinuous range of [$\alpha$/Fe] values. This has become known as the $\alpha$ bi-modality, due to its bi-modal distribution and a distinct ridge-line in [$\alpha$/Fe] vs. [Fe/H] space. Investigations have led to further discoveries related to the $\alpha$ bi-modality, such as the relationship between $\alpha$-richness and the geometrically and kinematically defined thin and thick disks \citep[e.g.,][]{GilmoreReid1983,bovy_rix_hogg2012,Hayden2015}.

%For instance, \cite{Hayden2015}---using $\sim$7,000 red giants from APOGEE DR12 \citep{Alam2015}---found that the $\alpha$-rich part of this sequence presents most significantly at scale heights above the Galactic plane $|Z| > 0.5$ kpc and within a Galactic radius of $R < 11$ kpc.

% Formation theories 
%As $alpha$-elements are only produced in SNe II it has to be true that the formation of the Milky Way is tied to supernova. However, the complex observed abundance pattern in $\alpha$-Fe demonstrates that there are other formation mechanisms at work. Other work has suggested other mechanisms that may contribute to the observed spread in abundance, including, the radial migration of stars in the Galaxy \citep{sellwoodbinney2002, Schonrich2009, Nidever2014, Weinberg+2017, sharma+2020}, two separate star formation episodes driven by the infall of gas into the Galaxy \citep{Chiappini1997, Spitoni2019, lian+2020}, and stars forming in clumpy bursts throughout the Galaxy \citep{Clarke2019}. Each of these models does a good job of recreating the observed pattern, however each create a difference prediction of the relative ages of $\alpha$-rich and -poor populations and their homogeniety across the Galaxy. Therefore, the importance of obtaining accurate ages for a wealth of stars in various populations cannot be understated. 
The literature presents several Galactic chemical evolution models that attempt to explain the observed spatial, chemical, and age trends associated with the $\alpha$ bi-modality. One class of models explains the $\alpha$ bi-modality with an initial, rapid star formation episode that forms the $\alpha$-rich population, with a subsequent lull in star formation that is pierced by an infall of pristine gas. This resets the metallicity of the disk, and quiescent formation proceeds to form the $\alpha$-poor population ("two-infall"; e.g., \citealt{Chiappini1997, Spitoni2019}). Another class of models describes two separate star formation episodes for the $\alpha$-rich disk and the $\alpha$-poor disk: the inner disk forms an $\alpha$-rich population at early times, followed by a smooth transition to an $\alpha$-poor population, with the inner part of the $\alpha$-poor population having higher metallicity than the outer part due to forming from the gas enriched by the inner disk \citep{haywood+2013,ciuca+2021}. An alternate scenario described by \cite{schonrich_binney2009} and expanded by \cite{sharma+2020} envisions a disk whereby stars occupying a wide range in chemical space are born simultaneously, though at higher rates in the inner disk than the outer disk. This causes $\alpha$-poor stars born in the slow chemical enrichment environment of the outer disk to have low metallicities that are otherwise associated with $\alpha$-rich stars. Radial migration \citep[e.g.,][]{sellwoodbinney2002} then brings populations of different chemistry into the solar neighborhood, causing the observed $\alpha$ bi-modality. A more recent model \citep{Clarke2019} predicts overlap in ages between $\alpha$-rich and $\alpha$-poor populations thanks to a clumpy star formation scenario, where star formation proceeds at different rates simultaneously throughout the disk in small clumps.

% READ , lian+2020 and place it in the above appropriately.

Although models have been shown to reproduce observed abundance and spatial trends, there has been little direct comparison between these model predictions and observed age-abundance patterns beyond generic predictions that the $\alpha$-rich population should be generally older than the $\alpha$-poor population. The consideration of the age trends in [Fe/H]-$\alpha$ space is therefore a potentially crucial test of these models, which we investigate here using comparisons between asteroseismic ages and the Galactic chemical evolution model of \cite{modelsource1}. 

In this paper, we expand on the work presented by \cite{warfield+2021}, where asteroseismic-based ages were derived for 735 RGB stars across three \ktwo\ campaigns. Here, we present accurate age measurements for 4,661 RGB stars from the APO-K2 Catalog (\citealt{catalogpaper}, hereafter \akccite), a cross-match between the asteroseismic \ktwo\ Galactic Archaeology Program (K2 GAP) catalog \citep{stello+2015, stello+2017} and APOGEE DR17. Thanks to the observing strategy forced upon the \ktwo\ mission, this catalog provides asteroseismic and spectroscopic parameters along 17 lines of sight in the MW, making it one of the most comprehensive asteroseismic-spectroscopic surveys of the diverse populations of our Galaxy to date.

% Paper breakdown 
In \S\ref{sec:data}, we discuss our data selection, showcasing the spectroscopic data in \S\ref{ssec:spec_data}, the asteroseismic data in \S\ref{ssec:astero_data}, and the selection function in \S\ref{ssec:selfun}. In \S\ref{ssec:cuts}, we discuss the cuts we have made to the APO-K2 data set. In \S\ref{sec:ages}, we detail our methods (\S\ref{ssec:method}), our comparison with the APOKASC-3 methodology (\S\ref{ssec:kepmethod}), and the effects of $\alpha$-abundance on age determination (\S\ref{ssec:alpha}). We compare our ages with spectroscopic ages from {\tt AstroNN} \citep{Mackereth+2019} in \S\ref{ssec:astronn}. We move on to analysing our population in \S\ref{sec:analysis}, with a discussion of the K2 fields in \S\ref{ssec:k2dist}, similarities between the Kepler field and the K2 fields in \S\ref{ssec:k2vkep}, stellar age and chemistry as a function of Galactic position in \S\ref{ssec:gal_pos}, and a comparison to modeled populations from \cite{modelsource1} in \S\ref{ssec:models}. We conclude and discuss our results in-context in \S\ref{sec:conclusions}. In addition to the online journal, age data is publicly available for the APO-K2 catalog at \url{https://github.com/jesstella/apo-k2} and \url{https://github.com/jackwarfield/apo-k2}.

\section{Data} \label{sec:data}

Our base data set is the APO-K2 Catalog (\akccite),\footnote{\url{https://github.com/Jesstella/APO-K2}} a cross-match between data from the \ktwo\ Galactic Archaeology Program Data Release 3 (K2 GAP DR3; \citealt{zinn+2022}), the Apache Point Observatory Galactic Evolution Experiment Data Release 17 \citep[APOGEE DR17;][]{majewski2017, SDSSdr17}, and \gaia\ DR3 \citep{gaia1, gaia3}. {In addition to this summary of the data, we detail the entire data pipeline, from K2 light curves to ages, in Appendix \ref{app:data}.}

\subsection{Spectroscopic Data} \label{ssec:spec_data}
APOGEE DR17 is a part of the final data release of the Sloan Digital Sky Survey Phase IV \citep[SDSS-IV;][]{SDSSIVoverview} and provides high-resolution near-infrared spectra (using twin, $R \sim 22{,}500$ $H$-band spectrographs; \citealt{wilson2019}) for 657,000 unique targets (with targeting described by \citealt{Beaton+2021}), encompassing observations dating back to the first edition of the APOGEE survey during SDSS Phase III in 2011. Spectra were collected using the 2.5-meter Sloan Foundation Telescope \citep{gunn2006} at the Apache Point Observatory in New Mexico, USA (APOGEE-North) from 2011-2020 and the 2.5-meter du Pont Telescope \citep{bowen73} at Las Campanas Observatory in Chile (APOGEE-South) from 2017-2020.

In addition to providing the raw data, APOGEE spectra have been put through a reduction pipeline that, in the end, provides spectrally-derived and calibrated estimates for stellar parameters such as effective temperature ($T_{\rm eff}$), log surface gravity ($\log{g}$), and chemical abundances (including ${\rm [Fe/H]}$ and ${\rm [\alpha/M]}$\footnote{${\rm [\alpha/M]}$ is conceptually equivalent to ${\rm [\alpha/Fe]}$, but instead measuring the ratio of $\alpha$-elements to the total metallicity (M) rather than just to Fe.}). \citeauthor{Nidever2015}
(\citeyear{Nidever2015}; with updates from \citealt{holtzman2018} and \citealt{jonsson2020}) describe the {schema} for extracting the spectra and performing wavelength calibrations, flat-fielding, and measuring radial velocities. % reads nicely. lots of juicy details like APOGEE-North v. APOGEE-South observing dates, which I've not seen written down

%\jtw{k2gap summary}
\subsection{Asteroseismic Data} \label{ssec:astero_data}
K2 GAP \citep{stello+2015} is the source for the asteroseismic data that we use in this work. The targeting for the program used simple color and magnitude cuts to select a sample of red giant solar-like oscillators, prioritizing bright and red targets. Dwarfs, which would not oscillate at low enough frequencies to be detected with \ktwo\ long-cadence data, were additionally selected against using a reduced proper motion selection cut. Details of the selection function can be found in \cite{sharma+2022}. The result is a well-understood sample ideal for Galactic archaeology applications.

\cite{zinn+2022} derived values for the asteroseismic parameters \numax\ and \dnu. These derived values for each star are made from the amalgamation of values from six independent pipelines, each based on an independent analysis of K2 light-curves, corrected for instrumental systematics \citep{luger+2018}.
We have further corrected \dnu\ using the prescription from \citealt{sharma+2016} in combination with APOGEE DR17 abundances (see \akccite{, and/or Appendix \ref{app:data:cal1},} for details). Stellar surface gravities ($g$, or $\log{g}$), masses, and radii are derived using asteroseismic scaling relations, which are calibrated to be on the \gaia\ DR2 radius scale (see \citealt{zinn+2022} {and/or Appendix \ref{app:data:cal1}}).
%\fillin{rc has no dnu correction, rgb reclassified as rgb younger. apok2 catalog paper describes}
 
%\cite{zinn+2022} derived values for the asteroseismic parameters \numax\ (the frequency of maximum acoustic power of the target's oscillations) and \dnu\ (the mean frequency spacing between the overtone modes of the oscillations). These derived values for each star are made from the amalgamation of values from six independent pipelines based on their independent analysis of \ktwo\ light-curves corrected for instrumental systematics \citep{luger+2018}. Given that \dnu\ is related to a star's mean density and \numax\ to the acoustic cutoff frequency (which, itself, is related to the star's surface gravity), it is possible to infer surface gravities ($g$, or $\log{g}$), masses, and radii for stars from these parameters through scaling relations linked to solar values \citep{tassoul1980, brown1991, kjeldsenbedding1995}. The masses and radii derived through these scaling relations were calibrated to be on the \gaia\ DR2 radius scale (see \citealt{zinn+2022}).
% \jcz removed chaplin+ 2008 reference.

\subsection{Selection Function} \label{ssec:selfun}
Because the targeting strategy used for selecting potential solar-like oscillators for K2 GAP is distinct from that used for targeting the same fields for APOGEE (which is described by \citealt{Beaton+2021}), the composite age of a given stellar population is potentially vulnerable to bias.
The selection function between these two targeting strategies has been worked out, in part, by \akccite\ {(see also Appendix \ref{app:data:selfun})} as functions of magnitude, color, mass, radius, \numax, and metallicity. Additionally, \akccite\ presents a selection function for translating from \ktwogap's stellar distribution to the true Galactic stellar populations of detectable asteroseismic giants as indicated by {\tt Galaxia} \citep{sharma+2011}. % \jcz the SF does not map to an underlying total stellar population

In the analysis of our results in \S\ref{ssec:k2vkep}, we present the composite ages of our sample's populations both un-scaled as well as scaled to the selection function in mass and metallicity that exists between the APO-K2 sample and the {\tt Galaxia} model; i.e., this selection function allows us to re-scale the APO-K2 age distributions to how they should approximately have appeared if the sample truly represented an unbiased sample of the Galaxy's asteroseismically detectable red giant population. The wider implications of accounting for selection functions for population age-dating is discussed in that section. % \jcz this is well put. it is a fine point re: what the selection functions actually do, which is not to correct for the true underlying population of all stars, just of the detectable asteroseismic giants.

\subsection{Cuts to the APO-K2 data set} \label{ssec:cuts}
The APO-K2 Catalog provides asteroseismically-derived masses and spectroscopic measurements of metallicity ([Fe/H]) and $\alpha$-element abundances ([$\alpha$/M]) for 7,672 red giant branch (RGB) and red clump (RC) stars. Stellar evolutionary states are assigned using a spectroscopic classification that has been calibrated using stars from the APOKASC-3 sample (M. Pinsonneault et al. 2024, in preparation), for which evolutionary states have been determined asteroseismically (our process is described in \citetalias{warfield+2021} using APOKASC-2 data, with updated parameters using APOKASC-3 % \jcz why Warf as cite alias haha
provided in \S2.3 of \akccite\ {as well as Appendix \ref{app:data:selfun}}). In this work, we derive stellar ages from stellar evolutionary tracks, using mass as a fundamental proxy for age. Therefore, we only consider stars for our analysis that are classified as being on the RGB. Though it is known that stars lose mass transitioning between the RGB and RC, without a detailed prescription of this change ages derived from the masses of RC stars will tend to be systematically biased to older ages (e.g., as shown by \citealt{casagrande+2016}; we discuss this further in Appendix \ref{app:rc}). We also limit the sample to stars with masses between 0.6 and 2.6 M$_{\odot}$, [$\alpha$/M] values between 0.0 and 0.4 dex, and [Fe/H] values between -1.0 and 0.6 dex, which is the parameter space encompassed by our evolutionary tracks and for which asteroseismic scaling relations are well behaved.\footnote{Though our tracks reach down to ${\rm [Fe/H]} = -2$, the behavior of the asteroseismic scaling relations where ${\rm [Fe/H]} \lesssim -1$ is still precarious (e.g., \protect\citealt{Epstein2014} and \protect\citealt{Valentini2019}). We look at calculating the ages for these stars in Appendix \ref{app:mp}.} % precarious is a nice word here---JS: agreed, love it!

\begin{figure}
    \includegraphics[width=\columnwidth]{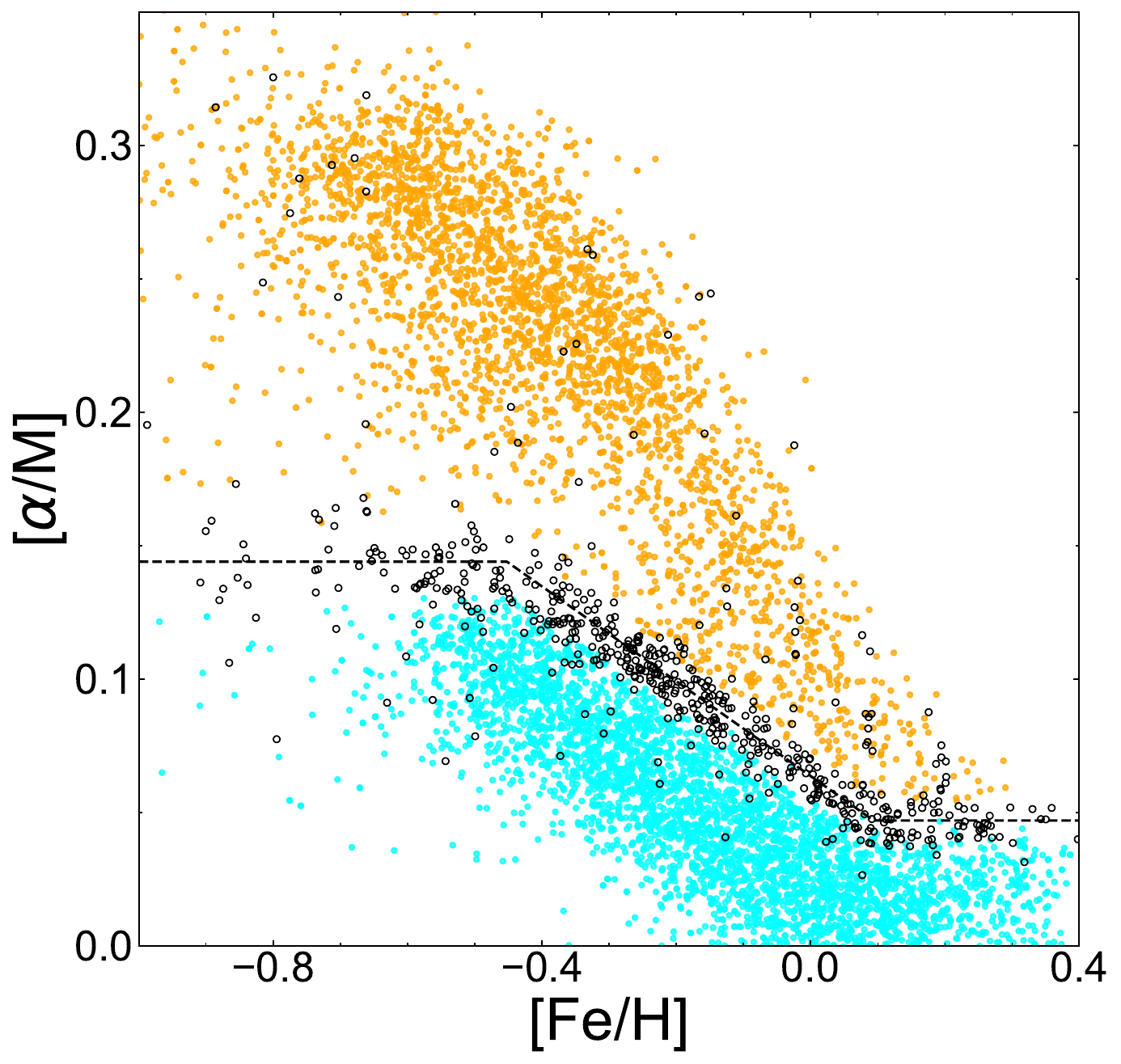}
    \caption{[$\alpha$/M] versus [Fe/H] for all stars in the \ktwo\ GAP sample. The black, dashed line represents our by-eye ridge-line to separate the $\alpha$-poor (cyan) from the $\alpha$-rich (orange) populations. The black unfilled markers represent stars that cannot be classified as either $\alpha$-poor or $\alpha$-rich with $>$95\% confidence, due to the uncertainties in their chemical abundances.}% nice description}
    \label{fig:alphaclass}
\end{figure}
In addition to our cuts, we have also defined two independent flags to categorize stars. The first is $\alpha$-rich versus $\alpha$-poor, which is made by drawing a ridge-line by-eye along the lower over-density of stars in the bi-modal [$\alpha$/M] vs. [Fe/H] distribution, with stars $\geq 2\sigma$ below the line being considered $\alpha$-poor (value of $0$ in the {\tt ALPHA\_RICH\_FLAG} column) and above the line as $\alpha$-rich ({\tt ALPHA\_RICH\_FLAG} = $1$). Stars falling within $2\sigma$ of this ridge-line are given neither classification ({\tt ALPHA\_RICH\_FLAG} = $-1$). Our $\alpha$-rich versus $\alpha$-poor classifications are shown in Figure~\ref{fig:alphaclass}.
The second classification is between the luminous and low-luminosity RGB (LL-RGB). Apropos asteroseismology, luminous giants are subject to measurement systematics due to the lower frequencies of their oscillations (versus their lower luminosity counterparts; \citealt{Mosser2013}, \citealt{APOKASC2}, \citealt{joelradius}). \citetalias{warfield+2021} show that the larger uncertainties of the luminous giants are able to moderately diffuse the distribution of ages for a given population of stars beyond its intrinsic spread. We explore this again in \S\ref{sec:analysis}, examining age distributions separately for stars with $\log{g} > 2.5$, which we classify as low-luminosity.
%\jss{Although we describe it in the APO-K2 catalog, I think it would be good to quickly go over how you do the evolutionary states, as it is so important for this work.}
%\jcz{I think the current description is fine --- with reference to previous paper where it is defined.}

\section{Age Determination and {Methodology Comparisons}} \label{sec:ages}
 
\subsection{Method} \label{ssec:method}
We have calculated underlying per-star age distributions in a manner identical to \S3 of \citetalias{warfield+2021}. Briefly, stellar evolutionary tracks generated with the {\tt Yale Rotating Evolution Code} ({\tt YREC}; \citealt{yrec1}, with updates from \citealt{yrec2}, and generated as described by \citealt{Tayar2017}) were used to create a regular grid (i.e., equally spaced along each axis), with axes for log(mass), [Fe/H], [$\alpha$/Fe], and log(age). Monte-Carlo sampling is used to draw sets of mass, [Fe/H], and [$\alpha$/M] values from each star's distributions (which are assumed to be Gaussian), and the implied age for each draw is then estimated from the grid of evolutionary tracks via multi-dimensional four-point Lagrange interpolation. {(See also Appendix \ref{app:data:ages}).} % nice 

\begin{splitdeluxetable*}{cccrrrrrrrBcrrrrrrrr}

\tablecaption{The partial data table for the APO-K2 RGB sample, including our ages. The complete table is available in CSV format in the online journal, as well as with the APO-K2 catalog online at \url{https://github.com/jesstella/apo-k2} and \url{https://github.com/jackwarfield/apo-k2}. The main identifier for each star is its ID in the Ecliptic Plane Input Catalog (EPIC). In addition to the columns shown here, our table contains Galactic longitude and latitudes; Galactocentric $R$ and $Z$ positions; the uncalibrated values for APOGEE $T_{\rm eff}$, $\log{g}$, [Fe/H], [$\alpha$/M], and [O/Fe], as well as the associated errors for all relevant columns. The {\tt $\alpha$-rich Flag} column has a value of {\tt 1} for $\alpha$-rich stars, {\tt 0} for $\alpha$-poor stars, and {\tt -1} for unclassified stars. The {\tt S.F. Weight} column is the weighting for that star given by the {\tt Galaxia} selection function, and have been normalized so that the maximum weight is 100. \label{tab:k2_table}}
\tabletypesize{\footnotesize}

\tablehead{
\colhead{EPIC ID} & 
\colhead{APOGEE ID} & 
\colhead{Gaia EDR3 Source ID} & 
\colhead{RA} & 
\colhead{DEC} & 
\colhead{$T_{\rm eff}$} & 
\colhead{$\log(g)_{\rm APO}$} & 
\colhead{[Fe/H]} & 
\colhead{[$\alpha$/M]} & 
\colhead{[O/Fe]} & 
\colhead{$\alpha$-rich Flag} & 
\colhead{Mass} & 
\colhead{Radius} & 
\colhead{$\log(g)_{\rm seis}$} & 
\colhead{$\nu_{\rm max}$} & 
\colhead{$\Delta\nu$} & 
\colhead{Age} & 
\colhead{Modal Age} &
\colhead{S.F. Weight} \\
\colhead{} & 
\colhead{} & 
\colhead{} & 
\colhead{(Deg.)} & 
\colhead{(Deg.)} & 
\colhead{(K)} & 
\colhead{($\log({\rm cm/s}^2)$)} & 
\colhead{(dex)} & 
\colhead{(dex)} & 
\colhead{(dex)} & 
\colhead{} & 
\colhead{(M$_\odot$)} & 
\colhead{(R$_\odot$)} & 
\colhead{($\log({\rm cm/s}^2)$)} & 
\colhead{($\mu$Hz)} & 
\colhead{($\mu$Hz)} & 
\colhead{(Gyr)} & 
\colhead{(Gyr)} &
\colhead{}
}

\startdata
220648976 & 2M01161528$+$1009159 & 2580092098586391168 & 19.0637 & 10.1544 & 4947 & 3.13 & -0.251 & 0.064 & 0.006 & 0 & 1.18 & 4.87 & 3.13 & 165.1 & 13.66 & 5.4 & 5.4 & 6.07 \\
212123262 & 2M08302828$+$2228487 & 665901492034489600 & 127.6179 & 22.4802 & 4509 & 2.46 & 0.203 & 0.025 & 0.048 & 0 & 1.35 & 10.32 & 2.54 & 44.0 & 4.73 & 4.6 & 4.6 & 55.68 \\
203757434 & 2M16095435$-$2502223 & 6049759992483427072 & 242.4765 & -25.0395 & 4487 & 1.91 & -0.663 & 0.312 & 0.371 & 1 & 0.60 & 12.47 & 2.03 & 13.5 & 2.38 & 45.7 & 30.0 & 0.51 \\
212570575 & 2M13283736$-$1113561 & 3611427412665830784 & 202.1557 & -11.2323 & 4797 & 3.06 & -0.270 & 0.226 & 0.260 & 1 & 1.00 & 4.65 & 3.10 & 156.8 & 13.51 & 10.6 & 10.6 & 8.25 \\
212458977 & 2M13271006$-$1336353 & 3609924895666745216 & 201.7919 & -13.6098 & 4783 & 2.90 & -0.378 & 0.265 & 0.357 & 1 & 0.97 & 5.46 & 2.95 & 110.1 & 10.44 & 11.3 & 11.3 & 5.94 \\
206005182 & 2M22072683$-$1440432 & 6827450163146087168 & 331.8618 & -14.6787 & 4702 & 3.19 & 0.289 & 0.060 & 0.077 & 1 & 1.16 & 8.86 & 2.61 & 50.4 & 5.52 & 8.5 & 8.0 & 70.08 \\
212562020 & 2M13490785$-$1124552 & 3613482601761697024 & 207.2827 & -11.4153 & 4874 & 3.09 & -0.364 & 0.274 & 0.295 & 1 & 1.08 & 4.47 & 3.17 & 180.3 & 14.82 & 7.7 & 7.1 & 5.16 \\
212396190 & 2M13564344$-$1500050 & 6301760184888854400 & 209.1810 & -15.0014 & 4787 & 2.85 & -0.439 & 0.290 & 0.324 & 1 & 0.92 & 5.64 & 2.90 & 98.2 & 9.70 & 13.1 & 12.7 & 4.70 \\
205976299 & 2M22254038$-$1531593 & 2596147343468963840 & 336.4183 & -15.5332 & 4989 & 3.14 & -0.661 & 0.283 & 0.247 & -1 & 1.39 & 5.42 & 3.11 & 157.1 & 12.65 & 2.5 & 2.5 & 0.18 \\
... & ... & ... & ... & ... & ... & ... & ... & ... & ... & ... & ... & ... & ... & ... & ... & ... & ... & ... \\
\enddata

\end{splitdeluxetable*}
In \citetalias{warfield+2021}, the median and $\pm 1\sigma$ percentiles of 500 log(age) values were reported for each star in the sample, and these values were used to construct the histograms and accompanying kernel-density estimations (KDEs) to discuss the overall characteristics of the populations. We repeat this process for our sample in this work, but have increased the number of runs from 500 to 5000 per star, made possible by optimizing the script from \citetalias{warfield+2021}. In addition to median ages, we have also calculated ages defined by the mode of a KDE that we have fit over the 5000 log(age) values for each star individually.\footnote{KDEs were fit using {\tt SciPy} (\url{https://scipy.org}; \citealt{scipy2020}).} This is done with the intention of spotting potential biases on the median from extended tails in the age distributions to unphysically old ages ($\gtrapprox$14 Gyrs). However, we use the median age estimates for all of our analysis in this work. Our ages for the APO-K2 Catalog sample are available in Table \ref{tab:k2_table} (as well as online at \url{https://github.com/jesstella/apo-k2} and \url{https://github.com/jackwarfield/apo-k2}).
%We also provide each star's KDE as functions of log(age) for download to allow readers to re-define their own statistic. \jtw{thinking about this, and how worth it it is actually to host these somewhere. Might be better to write just a small script that people can run on single stars to generate the KDE?} \jcz{that's a fine idea --- whatever is less work for you! hosting on zenodo would be the way to go.}

%In \citetalias{warfield+2021}, the median and $\pm 1\sigma$ percentiles of 500 log(age) values were reported for each star in the sample and these median values were used to construct the histograms and accompanying kernal-density estimations (KDEs) used to discuss the overall characteristics of the populations. In this work, we instead choose to fit each star's distribution of 5000 log(age) values with their own KDE and take the age at the peak (mode) and the $\pm1\sigma$ values around the peak as the age for each star. This is done with the intention of minimizing potential biases on the median from extended tails in the age distributions to unphysically old ages ($\gtrapprox$14 Gyrs). We also provide each star's KDE as functions of log(age) for download to allow readers to choose their own statistic. % so the differences are mode of the KDE and not median of the values + 5000 instead of 500 samples?

\subsection{Ages in the Kepler Field and Comparison with {an} APOKASC-3 Methodology} \label{ssec:kepmethod}
%\jcz{some intro words indicating that it is of interest to compare ages here with benchmarked Kepler ages}
\begin{splitdeluxetable*}{cccrrrrBrrcrrrrrrr}

\tablecaption{The partial data table for the re-calibrated APOKASC-2 RGB sample, including our ages. The complete table is available in CSV format in the online journal. The main identifier for each star is its ID in the Kepler Input Catalog (KIC), in the column {\tt KEPLER\_ID}. In addition to the columns shown here, our table contains Galactic longitude and latitudes; Galactocentric $R$ and $Z$ positions; the uncalibrated values for APOGEE $T_{\rm eff}$, $\log{g}$, [Fe/H] and [$\alpha$/M], as well as the associated errors for all relevant columns. The {\tt $\alpha$-rich Flag} column has a value of {\tt 1} for $\alpha$-rich stars, {\tt 0} for $\alpha$-poor stars, and {\tt -1} for unclassified stars. \label{tab:kep_table}}
\tabletypesize{\footnotesize}

\tablehead{
\colhead{Kepler ID} & 
\colhead{APOGEE ID} & 
\colhead{Gaia EDR3 Source ID} & 
\colhead{RA} & 
\colhead{DEC} & 
\colhead{$T_{\rm eff}$} & 
\colhead{$\log(g)_{\rm APO}$} & 
\colhead{[Fe/H]} & 
\colhead{[$\alpha$/M]} & 
\colhead{$\alpha$-rich Flag} & 
\colhead{Mass} & 
\colhead{Radius} & 
\colhead{$\log(g)_{\rm seis}$} & 
\colhead{$\nu_{\rm max}$} &
\colhead{$\Delta\nu$} &
\colhead{Age} & 
\colhead{Modal Age} \\
\colhead{} & 
\colhead{} & 
\colhead{} & 
\colhead{(Deg.)} & 
\colhead{(Deg.)} & 
\colhead{(K)} & 
\colhead{($\log({\rm cm/s}^2)$)} & 
\colhead{(dex)} & 
\colhead{(dex)} & 
\colhead{} & 
\colhead{(M$_\odot$)} & 
\colhead{(R$_\odot$)} & 
\colhead{($\log({\rm cm/s}^2)$)} & 
\colhead{($\mu$Hz)} & 
\colhead{($\mu$Hz)}  &
\colhead{(Gyr)} & 
\colhead{(Gyr)} 
}

\startdata
8176543 & 2M19414369$+$4405382 & 2079615472445407744 & 295.4321 & 44.0939 & 4366 & 2.06 & -0.124 & 0.060 & 0 & 1.10 & 17.08 & 2.02 & 13.4 & 1.93 & 7.8 & 7.3 \\
8277362 & 2M18440905$+$4417307 & 2117361186928151936 & 281.0377 & 44.2919 & 4507 & 2.40 & -0.300 & 0.260 & 1 & 0.96 & 10.64 & 2.37 & 29.4 & 3.65 & 12.6 & 12.7 \\
2161831 & 2M19270967$+$3731187 & 2051785390040144640 & 291.7903 & 37.5219 & 4902 & 3.05 & -0.330 & 0.122 & -1 & 1.06 & 4.94 & 3.08 & 145.0 & 12.41 & 7.6 & 7.4 \\
6664533 & 2M18452413$+$4209269 & 2104693683403661952 & 281.3506 & 42.1575 & 4567 & 2.70 & 0.172 & 0.058 & 1 & 1.16 & 7.98 & 2.70 & 62.8 & 6.25 & 8.1 & 8.0 \\
11723893 & 2M19473766$+$4949200 & 2087238180400724736 & 296.9069 & 49.8222 & 4651 & 2.80 & 0.135 & 0.035 & 0 & 1.15 & 6.83 & 2.83 & 84.1 & 7.88 & 8.0 & 7.9 \\
10517437 & 2M18524613$+$4745349 & 2107664151504239104 & 283.1922 & 47.7597 & 3895 & 1.48 & 0.234 & 0.019 & 0 & 1.28 & 36.22 & 1.43 & 3.7 & 0.68 & 6.5 & 6.0 \\
3441473 & 2M19232529$+$3833418 & 2052842158148906880 & 290.8554 & 38.5616 & 4643 & 2.57 & -0.346 & 0.295 & 1 & 0.99 & 7.84 & 2.65 & 55.3 & 5.90 & 11.0 & 10.9 \\
6501676 & 2M18552344$+$4159111 & 2104874759224452096 & 283.8477 & 41.9864 & 4609 & 2.56 & -0.184 & 0.117 & 1 & 1.04 & 9.14 & 2.53 & 43.0 & 4.81 & 9.1 & 8.9 \\
10272641 & 2M19250623$+$4718546 & 2129158263800815232 & 291.2760 & 47.3152 & 4859 & 2.86 & -0.223 & 0.063 & 0 & 1.29 & 6.95 & 2.87 & 89.8 & 8.17 & 3.9 & 3.9 \\
... & ... & ... & ... & ... & ... & ... & ... & ... & ... & ... & ... & ... & ... & ... & ... & ... \\
\enddata

\end{splitdeluxetable*}
Ages in the APOKASC-2 asteroseismic sample of stars in the \kepler\ field (as provided by \citetalias{APOKASC2}, but also by, e.g., \citealt{SilvaAguirre+2018} for APOKASC-1), being one of the largest homogeneous sets of accurate age estimates to-date, have become a pillar for investigating the evolutionary history of the  MW (e.g., \citealt{Spitoni2019}, {\citealt{Mackereth+2019}, \citealt{sharma+2021}, \citealt{sharma+2020}}). Expecting that the upcoming update to this sample will play a similar role (APOKASC-3; M. Pinsonneault et al. 2024, in preparation), it is crucial to understand how the \kepler\ and \ktwo\ samples compare, both from methodological and astrophysical standpoints.
Therefore, in addition to ages for the complete APO-K2 data set, we have calculated ages for the stars in the \kepler\ field, having recalculated the masses from \citetalias{APOKASC2} {with new} corrections to \dnu\ ($f_{\Delta\nu}$) {calculated} in the same manner as we have for APO-K2. As was done in \citetalias{APOKASC2}, our masses for stars in the \kepler\ field have been calibrated from an open cluster contained in the data set, which carries a $2.1\%$ systematic uncertainty due to uncertainties in eclipsing binary mass measurements. Our data for the Kepler Field can be found in Table \ref{tab:kep_table}.

As noted in \akccite, K2 GAP DR3 \numax\ values were calibrated to a different scale {than the APOKASC sample}: the \gaia\ DR2-based astrometric scale \citep{gaia2}, with a careful correction for parallax bias that took into account the different selection functions of RGB and RC stars \citep{schonrich+aumer2017, schonrichmcmillan+eyer2019, zinn+2022}. If we instead used parallaxes as corrected by the \gaia\ DR3 team without selection function corrections \citep{gaia2, lindegren+2021b} to perform the calibration, this would result in an $\sim 4.5\%$ downward revision of the mass scale. We acknowledge potential variations in the Gaia calibration by noting a $2\%$ systematic uncertainty in the \numax\ calibration scale and $6\%$ in mass \citep{joelradius} in the K2 data used in this work. Therefore, mass comparisons between K2 and Kepler {\it could} carry relative shifts with respect to each other up to a potential 6\% in mass, or 20\% in age. As we will see, it appears that the K2 and Kepler ages agree much better than this conservative estimate. {We also note that whatever offsets there might be in the native K2 and Kepler \numax\ scales (e.g., due to differences in the time baselines of the datasets; \citealt{sharma+2019,zinn+2022}), are removed by this Gaia calibration.}

\citetalias{warfield+2021} observed an offset of up to 2 Gyr between K2 stellar ages and those reported by \citetalias{APOKASC2}. The authors postulated that this could have potentially been due to the lack of $\alpha$-enhanced interior opacities in the models used by \citetalias{APOKASC2}, instead relying on the metallicity correction from \cite{salaris+1993}; we explore this hypothesis more fully in the following subsection (\S\ref{ssec:alpha}).

One of the APOKASC-3 age determination methods {will use} mass (with no assumption for RGB mass-loss), $\log{g}$, [Fe/H], and [$\alpha$/M] as look-up parameters in a {\tt YREC} stellar model grid {(also from }\citealt{Tayar2017}).
%APOKASC-3 will use multiple methodologies for calculating ages. {\color{red} One way ages will be calculated, differing from the methodology in APOKASC-2, is to use mass, $\log{g}$, [Fe/H], and [$\alpha$/M] as look-up parameters in the tracks generated by \cite{Tayar2017}}.
For the sake of comparison, we have generated ages {also} using this APOKASC-3 methodology for all of the RGB stars in the APO-K2 sample. As our methodologies rely on the same axes within a common set of tracks, but utilize slightly different 
%\jcz{I recognize this superior use of the word peculiar but for a non-native speaker (or for even native speakers), this may be read according to the less peculiar definition of the word peculiar so I would change to "different"}
look-up procedures, this comparison should provide a valuable sense of our method's systematic uncertainty.
\begin{figure}
    \includegraphics[width=\columnwidth]{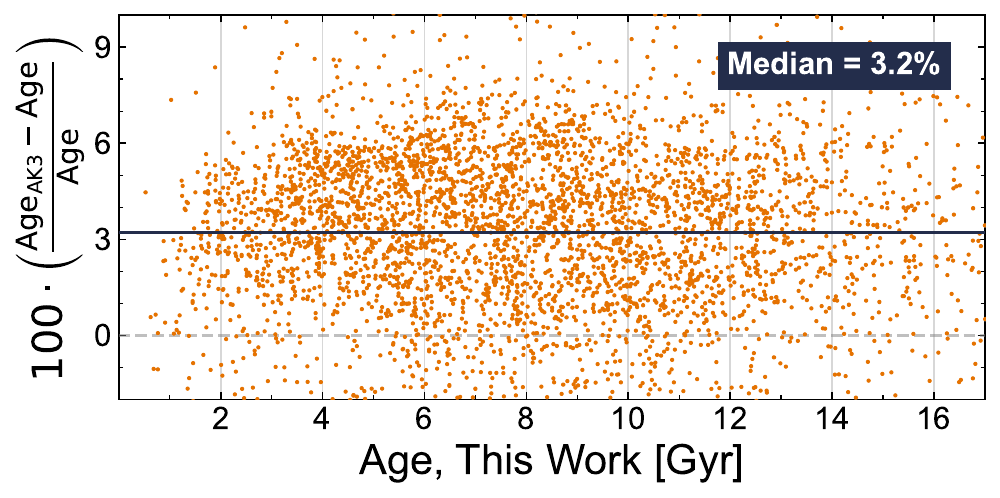}
    \caption{A comparison between the asteroseismic age determination method described in this work (and \protect\citealt{warfield+2021}) and a method to be used for the APOKASC-3 catalog (M. Pinsonneault et al. 2024, in preparation). The x-axis is ages from this work, in Gyr, and the y-axis is the percent-offset of the ages found using the APOKASC-3 methodology. The gray, dashed line represents the one-to-one line and the blue, solid line tracks the measured median offset between the two ages.} % I would put the median label farther up from the median line so we can see those points better
    \label{fig:ak3method}
\end{figure}
The comparison between these ages is shown in Figure~\ref{fig:ak3method}. In general, we see sufficient agreement between these ages, only with the APOKASC-3 methodology producing ages that are a median of 3.2\% older than ours, and consistently $\lesssim$6\% {older} than ours. Though not totally negligible on a star-by-star basis, this systematic offset is much smaller than the random uncertainties of our measurements.

One possible explanation for this zero-point offset is that it is due to how $\log{g}$ is taken into account by our different methods. In this work, we take the amount of time that a star spends on the main sequence (the main-sequence lifetime, or MSLT) as the age of a star.
The time for a star to move up the RGB and onto the asymmetric giant branch is $\lesssim10\%$ of its MSLT, and how far along on the RGB a star is will be associated with its $\log{g}$. Therefore, because this age-dating method from APOKASC-3 takes $\log{g}$ into account directly as a lookup parameter, it is {possible} to expect an average offset in ages at this level.
%(and will be even smaller for the lower uncertainties of the APOKASC-3 data). \jcz{Is this really true? Wouldn't it be more important because the uncertainties are smaller? I'm actually not sure how much smaller they even are, though, now that I think about it... best to leave this parenthetical remark out, methinks.}
Overall, this reminds us that age estimates from evolutionary tracks can be noticeably sensitive to methodology, even when working with the same sets of observational and theoretical data. This is especially relevant when taking into account that RGB ages derived from different stellar evolution codes also differ at the 2-5\% level, even for similar physical inputs \citep{rgbmodelageoffset}. %GP \jcz good reference

\subsection{Effects of Alpha-Abundance on Age Determination} \label{ssec:alpha}
\begin{figure*}
    \includegraphics[width=\textwidth]{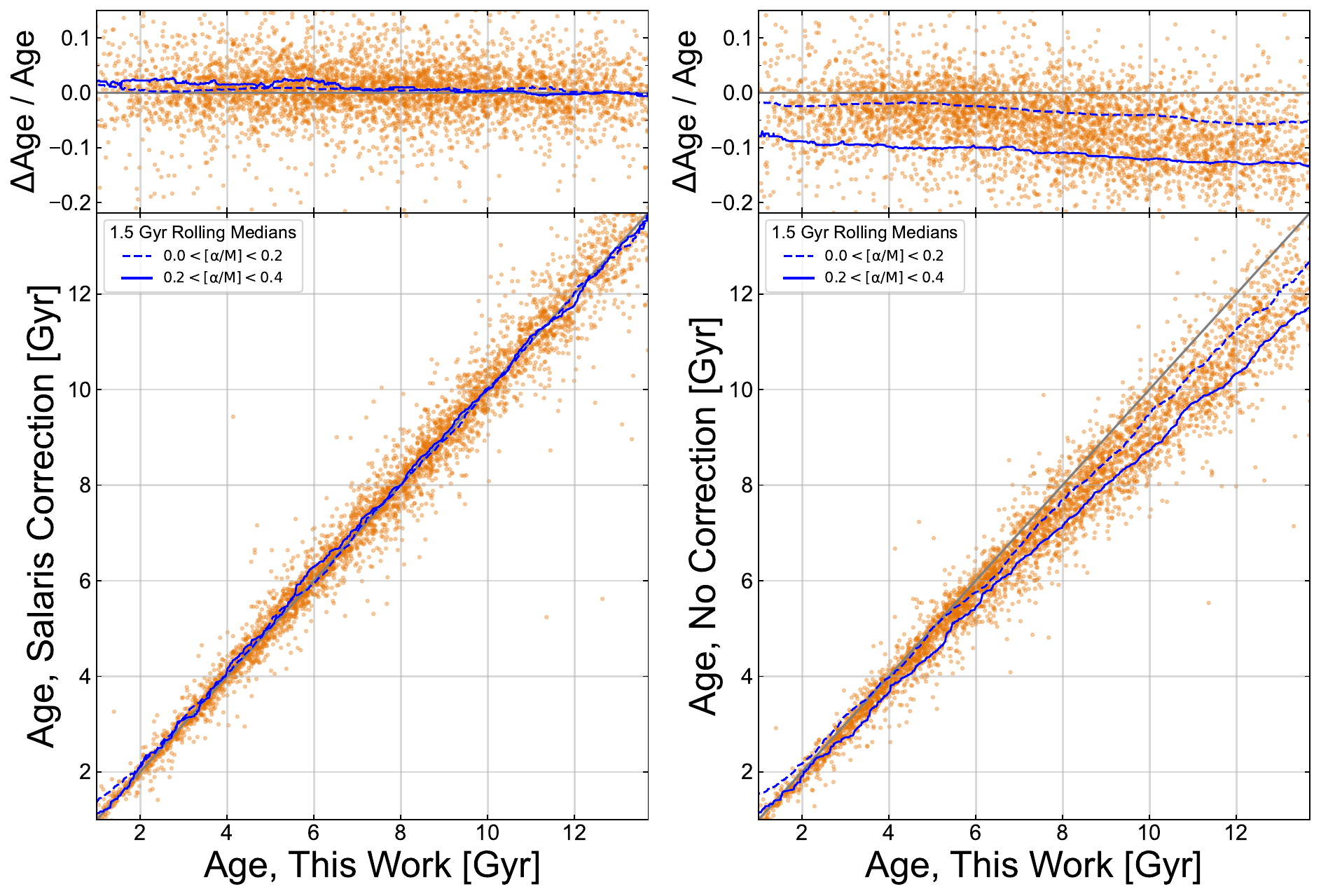}
    \caption{The left column of panels compares the ages for stars given in this work to the ages that would be obtained if interior opacity variations due to non-solar [$\alpha$/M] are not considered in the underlying evolutionary tracks, but rather each star's metallicity is corrected using the formula from \protect\cite{salaris+1993}.
    %The excellent agreement in this left column indicates that the \protect\cite{salaris+1993} approximation captures the effect of non-solar $\alpha$ abundances on age even using updated microphysics prescriptions.
    The right column has the same x-axis, but the y-axis shows the ages inferred if the metallicities of stars are given no correction.
    The bottom panels compare these ages one-to-one, and the top panels show the fractional offset of these ages versus ours. Each individual star in our data set is represented by an orange point, and we plot rolling medians with 1.5 Gyr bin-widths for stars with $0.0 < {\rm [\alpha/M]} < 0.2$ (blue, dashed line) and for $0.2 < {\rm [\alpha/M]} < 0.4$ (blue, solid line).
    The excellent agreement in this left column indicates that the \protect\cite{salaris+1993} approximation captures the effect of non-solar $\alpha$ abundances on age even using updated microphysics prescriptions.
    The large age disagreement shown in this right column demonstrates that the ages of $\alpha$-rich stars are strongly systematically biased by assuming solar $\alpha$ abundance. % this is a very nice illustration. salaris works but interestingly not 100% --- probably due to update of stellar physics. Do get rid of the x-axis ticks and labels for the upper plots to join the two columns into actual columns; place figure numbers for ease of reference; and consider plotting the fractional age difference and not the absolute age diff.
    }
    \label{fig:salaris}
\end{figure*}
In order to more fully explore the importance of $\alpha$-abundance in age calculations, we have estimated ages for each star in the APO-K2 sample with two alternative treatments of [$\alpha$/M]. These are:
\begin{itemize}
    \item[i.] ${\rm [Fe/H]}$ values are corrected using the prescription ${\rm [Fe/H]} = \log_{10}{(0.638 \cdot 10^{\rm [\alpha/M]} + 0.362)}$ from \cite{salaris+1993}, and then ${\rm [\alpha/M]}$ is set to a value of 0 when interpolating through the evolutionary tracks; and
    \item[ii.] ${\rm [\alpha/M]}$ is set to 0, with no correction applied to ${\rm [Fe/H]}$.
\end{itemize}
Calculating ages in the manner of (i) invites a test of the \cite{salaris+1993} correction against ages using models with non-solar $\alpha$ interior opacities taken into account, but with updated microphysics compared to those used in formulating the original correction (e.g., \citealt{grevesse+1998} abundance mixture; OPAL equation of state [\citealt{rogers_swenson_iglesias1996, rogers_nayfonov2002}]; and OPAL opacity tables [\citealt{iglesias_rogers1996}]). Calculating ages in the manner of (ii) quantifies the systematic age uncertainty incurred when not using the \cite{salaris+1993} correction at all.\footnote{We note that although $\alpha$-enhanced opacities are used in our models and in the \cite{salaris+1993} models, neither the equation of state tables used in our models nor those of the \cite{salaris+1993} models are calculated using $\alpha$-enhanced mixtures \citep{chieffi_streneiro1989}.} %\jcz have removed term 'self-consistent' for this reason.

The comparison between these two alternative treatments and our fiducial ages for the APO-K2 sample is shown in Figure~\ref{fig:salaris}. Along the left-hand column of this figure, we see the comparison between our catalog ages and the ages derived using (i). We can see that, as a median function of age, ages generated using the [Fe/H] correction from \cite{salaris+1993} are tightly consistent (no observed systematic offset) with ages generated when using [$\alpha$/M]-enhanced opacity tables in modern stellar structure calculations, with a scatter much tighter than the random uncertainty on these ages. In the right-hand column, we compare between our catalog ages and those generated using (ii). Here, we see a systematic offset to younger ages, at about the 2-5\% level for ${\rm [\alpha/M]} \leq 0.2$ and 10\% for ${\rm [\alpha/M]} > 0.2$.
%We also see, comparing the rolling medians for $\alpha$-poor stars versus the full sample, that the severity of this offset is a function of [$\alpha$/M]. 
This tells us two things:
\begin{enumerate}
    \item the prescription from \cite{salaris+1993} to take [$\alpha$/M] into account via a correction on [Fe/H] yields ages that are consistent to within $\sim 5\%$ with those from using $\alpha$-enhanced stellar models with updated microphysics;
    \item age is a non-negligible function of [$\alpha$/M] at fixed [Fe/H], and failing to properly account for $\alpha$-enrichment may lead to offsets of up to 2 Gyr for samples of old stars.
\end{enumerate}

%These results potentially run counter to the assumptions made in \cite{warfield+2021}, where it was assumed that the offset between their ages and the ages provided in the APOKASC-2 catalog at old ages was mainly due to the opacity effects of the different treatments of $\alpha$ by the evolutionary tracks used to calculate the respective ages. The conclusions above instead suggest that this explanation would only be the case if the ages given by \citetalias{APOKASC2} did not apply the correction from \cite{salaris+1993} to their metallicities.
These results somewhat muddy the {hypothesis} made in \citetalias{warfield+2021}, where it was assumed that the offset between their ages and the ages provided in the APOKASC-2 catalog at old ages was mainly due to the opacity effects of the different treatments of $\alpha$ by the evolutionary tracks used to calculate the respective ages. From our comparison above, it seems that, alternatively, a similar offset could be realized if the correction from \cite{salaris+1993} was never applied to the metallicities. However, as we discussed in \S\ref{ssec:kepmethod}, ages calculated with different sets of evolutionary tracks, or even merely differing look-up procedures, can lead to offsets at the $\sim$5\% level, and so it is still very possible that these various effects added together produced this offset.
%It is also possible that the offset seen in \cite{warfield+2021} is due to differences in model tracks. For instance, the tracks used for APOKASC-2 do not include a prescription for RGB mass loss, which could lead to offsets in inferred age \citep[e.g.,][]{casagrande+2016}. \jcz{important question here: do jamie's tracks have mass loss? if not this is a moot point. whatever they use, this should be mentioned when describing the models earlier.} \jtw{based on what she's said, I wouldn't think so?} \jnt{nope no mass loss in these tracks} % there is no mass loss in the APOKASC-2 age grid. \jtw{mass loss on rgb from aldo didn't greatly change it, may be important}
% Jamie: My tracks don't have mass loss, Aldo's have a bit of mass loss. But most of these stars are low enough on the RGB that not much mass has been loss yet- the loss really mostly matters for the clump stars. I suspect the differences between my ages and aldo's ages have more to do with different choices about overshoot, diffusion, abundance scales, etc. 

% We still note, however, that this is still likely only valid under the condition that the full effect of interior $\alpha$-abundance on opacities has been taken into account in the stellar models. However, it may be a valid choice to rely on this correction in order to reduce the dimensionality of the age lookup, for instance. In fact, it is possible that this will give more accurate results when interpolating within tracks that are more finally sampled in [Fe/H] than [$\alpha$/Fe].

\subsection{Comparison with spectroscopic ages from {\tt AstroNN}} \label{ssec:astronn}

{\tt AstroNN} is an APOGEE value-added catalog that provides stellar abundances, distances \citep{Leung+Bovy2019}, and ages \citep{Mackereth+2019} for stars through implementations of Bayesian convolutional neural networks. In particular, hoping to exploit the relation between surface C and N abundances and stellar mass/age in red giants (e.g., as presented by \citealt{Salaris+2015}, and also explored in the APOGEE data by \citealt{Martig+2016a}), \cite{Mackereth+2019} derives ages for red giants in the APOGEE catalog using ages from APOKASC-2 and the associated APOGEE elemental abundances (since updated using APOGEE DR17) as the training set. % nice summary!
% \jcz also nice reference to salaris+2015 --- wasn't aware of the theoretical work done before martig.

\begin{figure}
    \includegraphics[width=\columnwidth]{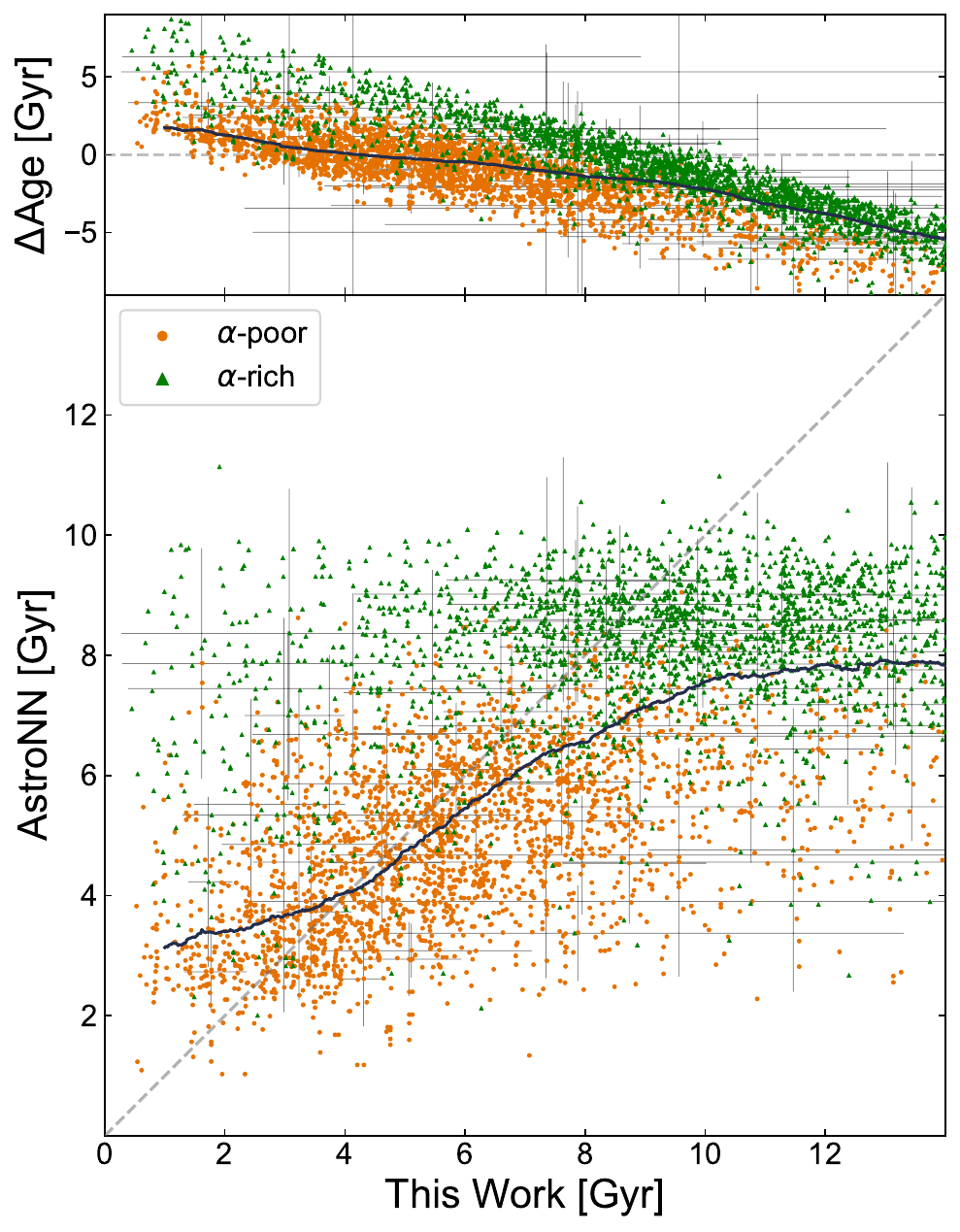}
    \caption{A comparison between age estimates for RGB stars in the APO-K2 sample produced using the method described in this work and \protect\cite{warfield+2021} versus those produced by \protect\cite{Mackereth+2019} with {\tt AstroNN}. The bottom panel compares these ages directly and the top panel shows the age offsets, $\Delta{\rm age} = \tau_{\rm NN} - \tau$. The gray, dashed line represents the one-to-one line and the blue, solid curve tracks the rolling inverse-variance weighted-mean of the data with a bin width of 2~Gyr. Orange, circular markers are stars classified as $\alpha$-poor and green, triangular markers are for $\alpha$-rich stars. In order to not overcrowd the plots, error bars are shown for a random 1\% of the sample.}
    \label{fig:astroNN}
\end{figure}
The comparison between our ages and the ages for the same RGB stars from \cite{Mackereth+2019} is shown in Figure~\ref{fig:astroNN}. For our sample, age versus age trends are similar to what is shown between {\tt astroNN} and APOKASC-2 in Appendix A of \cite{Mackereth+2019}. We find that, below $\sim$4 Gyr, {\tt astroNN} tends to predict ages up to $\sim$2 Gyr larger than ours, and above $\sim$8 Gyr, {\tt astroNN} tends to predict lower ages, with an offset of up to 5~Gyr for the oldest stars. The {\tt astroNN} ages also demonstrate a clear upper limit of approximately 10 Gyr. Similar trends have been found in other works independently deriving spectroscopic masses and ages (\citealt{Martig+2016a}; \citealt{das_sanders2019}; \citealt{Ting&Rix2019}; \citealt{anders+2023}; \citealt{wang+2023}; \citealt{stone-martinez+2023}), though encouraging efforts have shown progress in addressing bias in spectroscopic ages (e.g., \citealt{ciuca+2021}; \citealt{leung+2023}).

Regarding the mismatch at old ages, the $\alpha$-rich population is known to have a very strongly peaked age distribution, which is located at $\sim$9 Gyr for the APOKASC sample when using the ages provided by \citetalias{APOKASC2}.
[C/N] is not strongly mass-dependent at low masses (see, e.g., \citealt{Martig+2016a}, Figure~3; {\citealt{Roberts+2024}}). Therefore, algorithms relying on [C/N] are likely to assign any low-mass star to approximately the typical mass/age of an $\alpha$-rich star within random variation.

However, in general, this limitation for spectroscopic methods at old ages may not be as problematic as it initially appears. As it stands, {\tt astroNN} is quite effective at tagging low-mass/old stars. \citetalias{warfield+2021} (as well as, e.g., \citealt{SilvaAguirre+2018}) show that the spread in ages around the median for $\alpha$-rich stars in the Kepler field is consistent with those stars' random age uncertainties (which we show to be the case for the K2 fields in \S\ref{ssec:k2vkep}). Similarly, the standard deviation for the $\alpha$-rich {\tt astroNN} ages in Figure~\ref{fig:astroNN} is comparable to the spread in the APOKASC-2 $\alpha$-rich ages. The mismatch in the Figure~\ref{fig:astroNN} $\alpha$-rich age spread and that of K2 is driven by the larger K2 asteroseimic age uncertainties compared to APOKASC-2 {(\S\ref{ssec:k2vkep})}. Therefore, if it is true that the $\alpha$-rich population is approximately coeval across the Galaxy, and the APOKASC sample is the most precise measurement of this population's age, then {\tt astroNN} would be accurately predicting the age of these stars at the composite level.
However, this does still rely heavily on the assumption of a coeval $\alpha$-rich population. If there are genuine, intrinsic, astrophysical trends at old ages, or a significant position--age relation at old ages, for instance, then a spectroscopic method, assigning a median $\alpha$-rich age, would be unable to uncover them.

At young ages, {\tt astroNN} and other spectropic age determinations face the same limitations as the asteroseismic RGB data. That is, though the trends found from studying RGB ages may still remain (adjusted by a multiplicative zero-point, and perhaps with more random scatter than the underlying asteroseismic data set), there are still important questions to be answered regarding how well the age trends derived from RGB stars actually are reflective of, and can be applied to, the general underlying stellar population. For instance, because a star's RGB lifetime is proportional to its MSLT, we would expect to find few young RGB stars. RC stars, with their longer lifetimes, may be a valuable tracer for the true density of younger stellar populations. {\cite{Roberts+2024}} does show RC stars to have a more reliable [C/N]--mass relationship at high mass, potentially offering a promising avenue for tackling this question. However, today, RC stars still have limited utility, as it is difficult to put these stars on an absolute age scale due to their yet-fully understood mass loss (see Appendix \ref{app:rc}).

\section{Population Analysis} \label{sec:analysis}

\subsection{K2 age distributions} \label{ssec:k2dist} %\label{ssec:k2vkep}
\begin{figure*}
    \includegraphics[width=\textwidth]{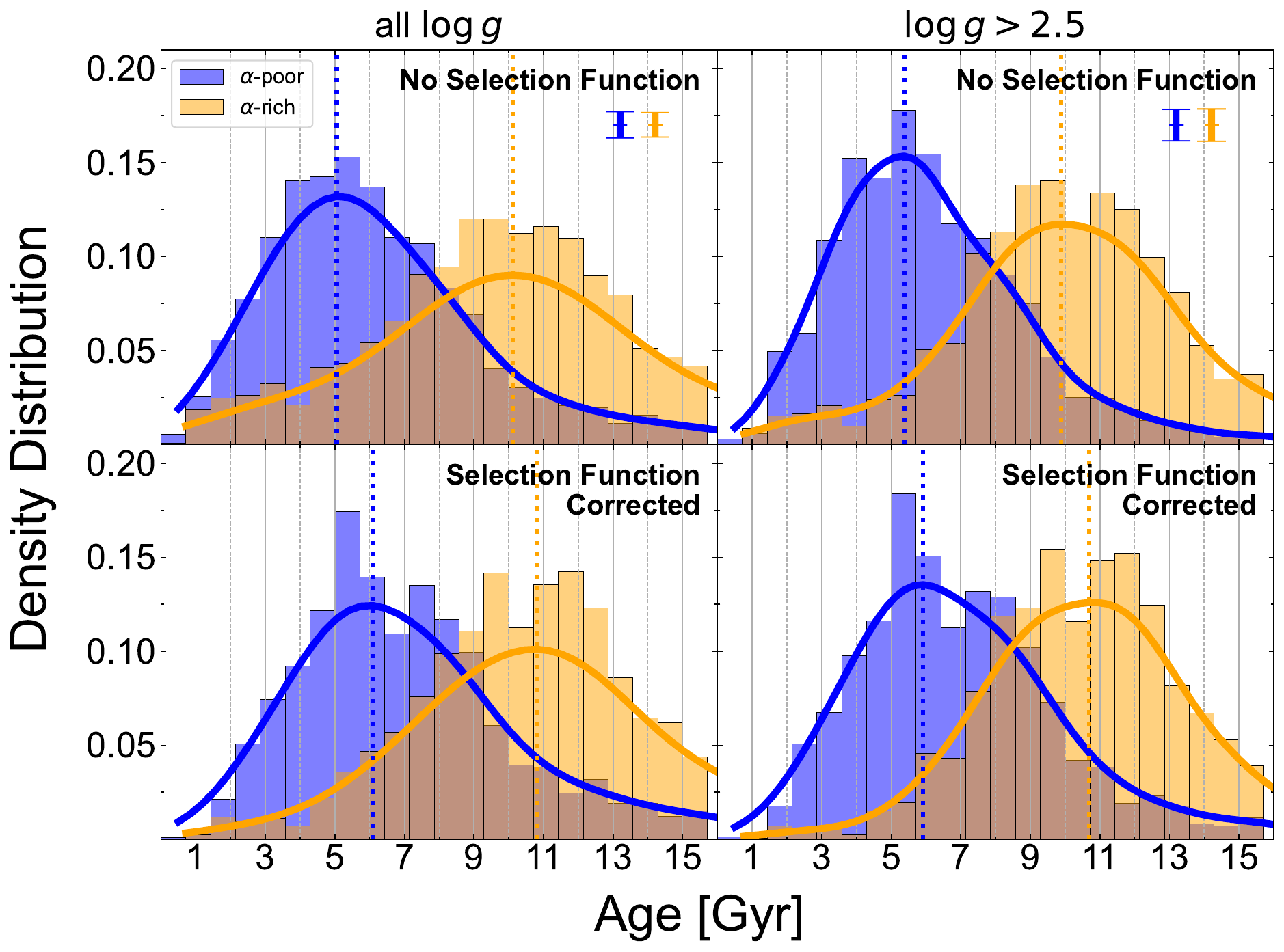}
    \caption{Density distributions for the ages of stars in the $\alpha$-poor (blue) and $\alpha$-rich (orange) RGB populations across all \ktwo\ campaigns. Densities are calculated per chemical population, such that the areas under the $\alpha$-poor and $\alpha$-rich curves histograms both, individually, sum to one, and so the relative heights of these distributions are not reflective of the relative stellar counts between each population. Along the top row, we plot these distributions for our unaltered sample and, on the bottom, for our re-scaled sample, accounting for the selection function from \protect\akccite. The vertical, colored dashed lines mark the modal peak of each distribution. In the top right of the top panels, we show the representative Poisson uncertainties at a density of 0.1 for both the $\alpha$-poor and $\alpha$-rich sample's histograms. The plots in the left column include stars within the full range of $\log{g}$, which includes 2176 $\alpha$-poor and 2467 $\alpha$-rich stars, where the right column is limited to the low-luminosity RGB ($\log{g} > 2.5$), with 1463 $\alpha$-poor and 1402 $\alpha$-rich stars. In all panels, only stars with ${\rm [Fe/H]} < 0.35$ are included, restricting the sample to the range within which our selection function can be interpolated.}
    \label{fig:k2composite}
\end{figure*}
For Figure~\ref{fig:k2composite}, we have plotted one-dimensional age histograms for the $\alpha$-poor and $\alpha$-rich populations across all K2 campaigns within our sample. In the top panel, we have included all stars for which $-1.0 {\rm\, dex} \leq {\rm [Fe/H]} \leq 0.35$ and $0.60 {\rm\, M_\odot} \leq M \leq 2.6 {\rm\, M_\odot}$,\footnote{These limits in metallicity and mass were chosen because it is the range in which the bins of the selection function are well sampled.} with no weighting, and in the bottom panel we have re-weighted these same stars using the {\tt Galaxia}-to-K2 metallicity and mass selection function (described in \S\ref{ssec:selfun}). In the left column of this plot we include stars with all values of $\log{g}$, where in the right column we only include stars on the LL-RGB (\S\ref{ssec:cuts}).

We see, in comparing the widths of the distributions in the first row of Figure~\ref{fig:k2composite}, that the smaller age uncertainties for the LL-RGB stars (with median age uncertainties of $\pm_{1.7}^{2.2}$ Gyrs, versus $\pm_{2.1}^{2.9}$ Gyrs for the full sample) result in correspondingly smaller spreads in the age distributions. This is perhaps more notably the case for the $\alpha$-rich population, where the standard deviation of ages for for the full sample is 5.6 Gyrs versus 4.2 Gyrs for the LL-RGB, and which is a population with considerably less intrinsic age spread versus the $\alpha$-poor population. Additionally, the percentage of very young $\alpha$-rich stars ($<4$~Gyr) in the LL-RGB sample is considerably lower (4.5\% versus 8.5\%). In the full catalog, there is a comparable number of stars at $<4$~Gyr as there are $>25$~Gyr. In the LL-RGB sample, there are more exceptionally young than exceptionally old stars, suggesting that the LL-RGB population of young $\alpha$-rich stars may be a purer sample of truly young {or otherwise} high-mass sources. However, despite these changes, the median ages of the $\alpha$-rich populations from these two samples are essentially the same (11.5 Gyrs for the full sample, 11.6 Gyrs for the LL-RGB). We suggest that this indicates that the LL-RGB represents a valid, precise subset of our sample with no biasing effect on the distributions of our populations defined by $\alpha$ abundance, and therefore we may expect the distributions from this subset to be more indicative of the intrinsic spread in ages for these populations. It also suggests that the population of young $\alpha$-rich stars may partially, but not entirely, be a product of age uncertainty (e.g., as discussed by \citealt{Anders2017}). % i have softened this language because they do show up in apokasc data and

Along the bottom row of Figure~\ref{fig:k2composite}, we see that the selection function weighted distributions are very similar to the unweighted distributions. Particular differences are that the {\tt Galaxia} models seem to predict slight shifts to older ages in both populations, mainly predicting fewer very young $\alpha$-poor stars, with a larger overlap between the two populations at intermediate ages. We also see more pronounced secondary modes in both of these populations. {In particular, the double-peaked profile of the $\alpha$-poor population with the selection function applied now more more closely the profile of the $\alpha$-poor population in the Kepler field (\S\ref{ssec:k2vkep}).} Otherwise, the median ages of both populations agree within uncertainties.

\subsection{Kepler age distributions and similarities with the K2 fields} \label{ssec:k2vkep}
\begin{figure*}
    \includegraphics[width=\textwidth]{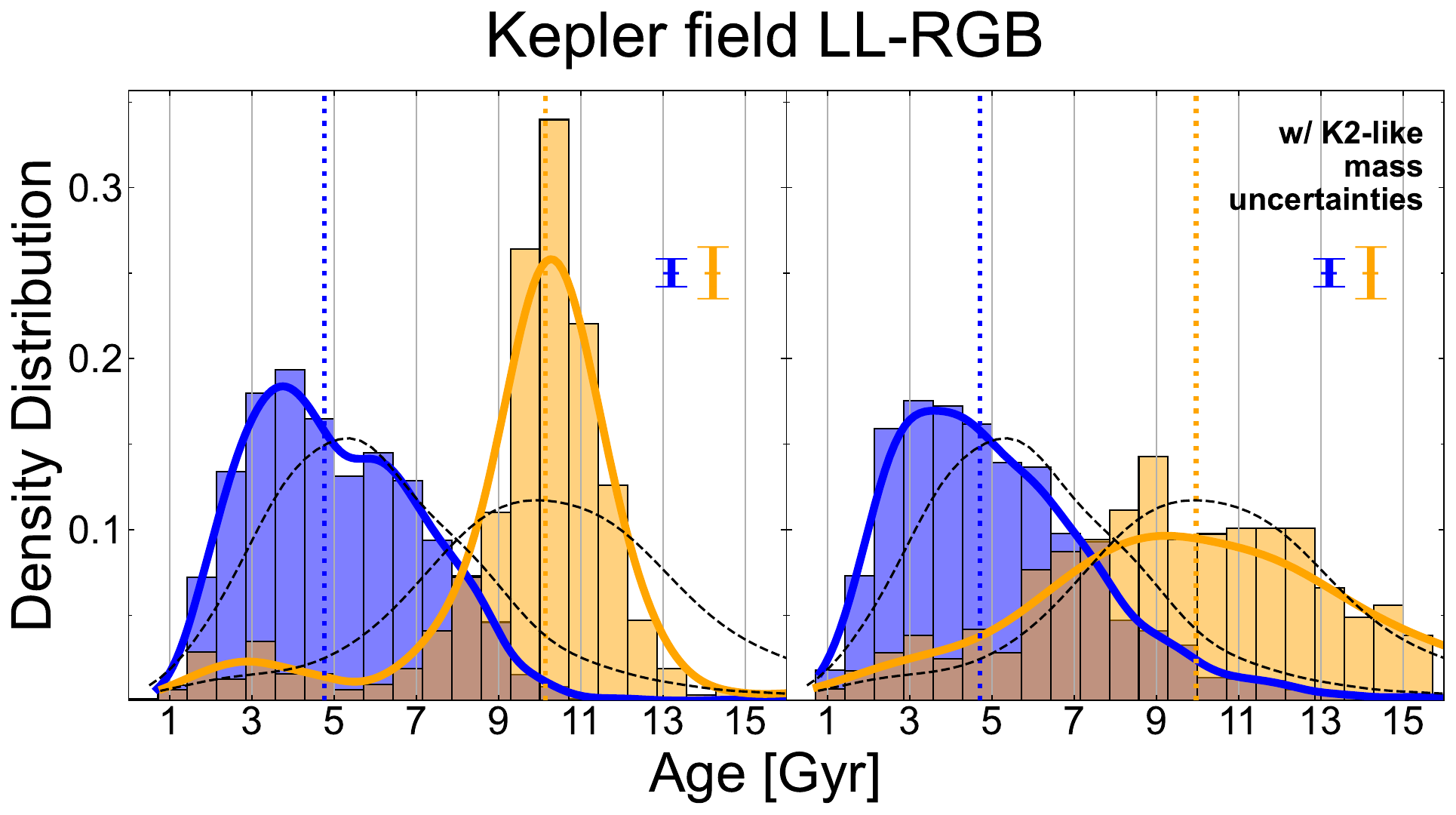}
    \caption{Density distributions for the ages of stars in the $\alpha$-poor (blue; 1557 stars) and $\alpha$-rich (orange; 452 stars) LL-RGB populations within the Kepler field. Densities are calculated per chemical population, such that the areas under the $\alpha$-poor and $\alpha$-rich histograms both, individually, sum to one, and so the relative heights of these distributions are not reflective of the relative stellar counts between each population. On the left, we plot the age distribution as it would be calculated using the methodology described in \S\ref{ssec:method}.
    On the right, we show the ages for the same stars calculated after those stars' masses have been randomly perturbed randomly by the typical mass uncertainties for similar stars in the K2 fields.
    The dashed curves on each plot represent the KDE fits for the K2 LL-RGB sample, which is also shown in Figure~\protect\ref{fig:k2composite}. In the top right of each panel, we show the representative Poisson uncertainties at a density of 0.1 for both the $\alpha$-poor and $\alpha$-rich sample's histograms.}
    \label{fig:kepler}
\end{figure*}
In Figure~\ref{fig:kepler}, we present the age distribution for LL-RGB stars in the Kepler field. On the left, we show the ages for these stars, calculated as described in \S\ref{ssec:method} (with masses re-calibrated from APOKASC-2, as described in \S\ref{ssec:kepmethod}).
Overall, these age distributions show good agreement with those for the same sample of stars made by \cite{SilvaAguirre+2018}, with the primary peak of the $\alpha$-poor distribution coming in at $\sim3$~Gyr, and the $\alpha$-rich distribution being strongly peaked at $\sim10$~Gyr.

In comparison to the K2 sample, however, though the median ages of the $\alpha$-rich and $\alpha$-poor populations are consistent between the samples, there is still a qualitative difference {in} the shapes of these distributions.
On the right side of Figure~\ref{fig:kepler}, we have plotted the age distribution of the \kepler\ field as it would appear if the masses of those stars had the same uncertainties as similar stars in the APO-K2 data. We have also, on both sides of the figure, plotted the distributions of the APO-K2 LL-RGB $\alpha$-poor and $\alpha$-rich samples as dashed lines.
%However, as is also discussed in \S\ref{ssec:kepmethod}, these ages have been shifted in order to force the median ages of the {\it Kepler} and \ktwo\ $\alpha$-rich populations to agree. Given this shift
Comparing these scenarios, it seems that, by accounting for the differences in the asteroseismic mass (and, therefore, age) uncertainties between the samples, we can infer that the \kepler\ field's stronger $\alpha$-rich peak and the \ktwo\ fields' lack of a potentially double-peaked $\alpha$-poor age distribution (both seen by, e.g., \citealt{SilvaAguirre+2018} and \citealt{Miglio2021} in the Kepler field) may be due---at least in part---to the difference in data quality between the samples (as is also noted by, e.g., \citealt{rendle+2019}). Inversely, we also might infer that the $\alpha$-rich stars of K2 are consistent with the $\alpha$-rich population being approximately coeval as, with K2-like uncertainties, the Kepler field's $\alpha$-rich distribution seems to match the qualitative features of the K2 distribution, including the appearance of a potential second peak.  % versus being differences in the innate properties of these fields.

Despite some differences, the existing agreement between the K2 and Kepler age distributions shows the reassuring progress made with these data since \citetalias{warfield+2021}. In \citetalias{warfield+2021}, the authors found a median age of $\sim8.6$~Gyrs for the $\alpha$-rich populations from Campaigns 4, 6, and 7 (K2 GAP DR2; \citealt{zinn+2020}), with this population having a modal age about 2~Gyrs lower than the age they found for the Kepler field. The non-astrophysical explanation given by the authors was that this was related to the overall \numax\ scaling used for K2 asteroseismic data. In K2 GAP DR2 and \citetalias{warfield+2021}, the scaling was not tied to Gaia, and so an $\sim$2\% systematic remained between the asteroseismic and astrometric radii. Using the K2 GAP DR3 data (which is calibrated to the Gaia radius scale), we now find the median age for the same set of $\alpha$-rich stars from \citetalias{warfield+2021} to be $\sim$20\% larger, at $\sim$10.4~Gyr, which is more consistent with the age of similar stars in the Kepler field.

As we will explore in the next section, remaining discrepancies between the K2 and Kepler populations may due to real differences between stellar populations{, tied to their spatial distributions} across the Galaxy. \cite{stokholm+2023} similarly support this idea with age data, showing the evolution of the ages of stellar populations as a function of Galactocentric radius. This idea is also supported by \cite{sharma+2022}. They compare the asteroseismic masses of the Kepler and K2 samples, {both observationally and theoretically (based on Galactic models)}, to show that there is potentially a real difference in the modes of stellar masses across different fields independent of spectroscopic data.

\subsection{Age versus Galactic position in the Kepler field and K2 fields} \label{ssec:gal_pos}
As we have stated above, the K2 fields are unique compared to the Kepler field because---apart from observing strategies and time baselines---they cover a much wider positional sample of the MW, both in terms of radial distance from the Galactic center ($R$) and vertical distance from the plane of the Galaxy's disk ($Z$).

Previous studies, such as \cite{Hayden2015} using APOGEE DR12 \citep{holtzman+2015}, have already shown that both the relative and absolute distributions of stars {in chemical phase-space} are functions of both $R$ and $|Z|$, with, {for instance}, more $\alpha$-rich stars appearing at larger $|Z|$, and mostly within $R\lesssim 11$\,kpc (representing the older, "thick" disk) and $\alpha$-poor stars being present at all values of $R$, but with $|Z| \lesssim 1$\,kpc. In addition to these trends in the number of stars in each population, \cite{Hayden2015} observed that $\alpha$-poor stars in the range $3{\rm\,kpc} < R < 5{\rm\,kpc}$ have typical [Fe/H] values of about $0.2$~dex, where stars on the outskirts of the Galaxy ($13{\rm\,kpc} < R < 15{\rm\,kpc}$) have a typical [Fe/H] of about $-0.4$~dex.

% \jcz i think this is a red herring: the ages that you are inferring are the real ages, no matter what the alpha and feh are. the reason that the K2 and Kepler age distributions could be skewed from the true, underlying thin and thick disc populations is a result of mostly this Z-effect, where older stars have had time to be puffed up (or they were born earlier at higher orbits). i think that is the real thing that should be discussed here, as well as the radial migration thing per comment later. As we explored in \S\ref{ssec:alpha}, the variations in the $\alpha$ abundances of two stars of the same mass have significant influence on the relative inferred age of those stars. I.e., though the relative counts of stars in each chemical population can be normalized in order to express the general distribution of stellar ages drawn from a given population (as we do in \S\ref{ssec:k2vkep}), the age distributions estimated for the $\alpha$-rich and $\alpha$-poor stars in a single field may not serve as an adequate description for the age distributions of these populations more broadly.

\begin{figure}
    \includegraphics[width=\columnwidth]{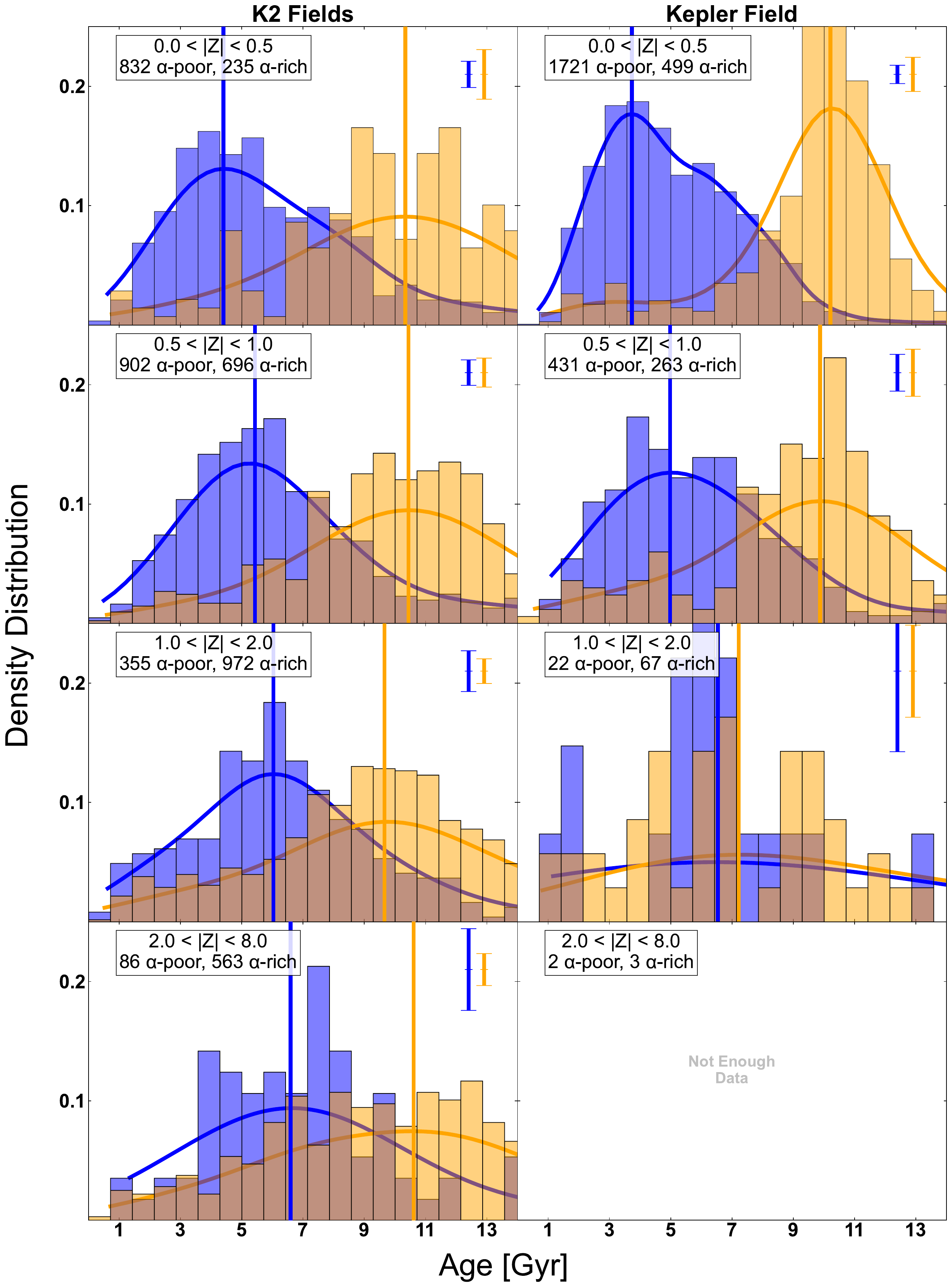}
    \caption{Density distributions for the ages of stars in the $\alpha$-poor (blue) and $\alpha$-rich (orange) RGB populations within the {\it K2} fields (left) and {\it Kepler} field (right), in different bins of height above or below the Galactic plane, $|Z|$. Densities are calculated per chemical population, such that the areas under the $\alpha$-poor and $\alpha$-rich curves both, individually, sum to one, and so the relative heights of these distributions are not reflective of the relative stellar counts in each population. In the top right corner of each panel, we show the representative Poisson uncertainties at a density of 0.1 for both the $\alpha$-poor and $\alpha$-rich samples histograms.} %\jcz{The relative age differences between $\alpha$-poor populations in K2 v. Kepler seen in \protect{\ref{fig:k2composite}} is seen here to be due to differences in the Galactic latitudinal distributions of the K2 and Kepler samples.}\jcz{I keep trying to trace the two populations of alpha-poor stars that are seen in Fig. 5 at 8 and 5 Gyr as well as the two high-alpha populations at 12 and 10 gyr. (there is also the bi-modal thing in poor stars for 4 and 6 gyr, but upon this read, the peak at 8 strikes me.) for the alpha-poor stars, i would use Zmax instead of |Z|, which you can get from jess's table. I am wondering if that 8Gyr peak is actually just stars that normally sit at larger Z but happen to be seen at the present Z now.  the alpha-rich situatino is more confusing because there is a wobble in the median age. I wonder if part of this is that some of the GE stars are getting in (the met. rich ones). So I would try cutting those out explicitly using Jess's kinematic cuts. can't do that as easily with the Kepler data because not having the kinematics handy (though i think they are in Jamie's APOKASC catalog). But this is just to see what happens. maybe it will clear them up. Jess's selection does not select that many of them; I would use these cuts instead: |Lz | < 0.5 (×103 kpc km s−1) ; –1.6 < E < –1.1 (×105 km^2/s^2) --- from horta paper https://ui.adsabs.harvard.edu/abs/2023MNRAS.520.5671H/abstract --- to clear up the low-alpha situation. It would also be informative to make a scatter plot of feh v. Zmax and highlight the following pops: 5gyr$\pm$ 0.5 poor (or try to separate out the 4 and 6 gyrs ones), 8 gyr$\pm$ 0.5 poor, 10$\pm$ rich, 12$\pm$ rich.}}
    \label{fig:agevz}
\end{figure}
Looking at both the K2 and Kepler fields, we do not see obvious trends in the age of the $\alpha$-poor population with $R$ alone. However, we do see a clear dependence on $|Z|$, which is shown in Figure~\ref{fig:agevz}.\footnote{Each star's Galactocentric $R$ and $Z$ position was calculated with {\tt Astropy} (\href{http://www.astropy.org}{http://www.astropy.org}; \citealt{astropy:2013}; \citealt{astropy:2018}; \citealt{astropy:2022}) and using Gaia eDR3-based distances calculated using the methodology described by \cite{BailerJones2015}.} Similar to what was shown by \citetalias{warfield+2021}, we see that that the peak age of the $\alpha$-poor population shifts older with increasing $|Z|$. In the K2 fields, this age is just above 4~Gyrs under 0.5~kpc, versus $\sim$6~Gyrs above 2~kpc. This seems to be connected to the bi-modal age distribution of the $\alpha$-poor population. This bi-modal distribution is most obvious in the Kepler data, where the majority of stars are at $|Z| < 0.5$~kpc{, and where} we observe peaks in the $\alpha$-poor distribution both around 3~Gyrs and 6~Gyrs. In fact, when we look at K2 stars with $|Z| < 0.5$~kpc (Figure~\ref{fig:agevz}, first panel), we see the younger peak emerge in this bi-modal distribution, which is not clearly present---nor dominant---when all of the data is combined. Instead, we see that, in the composite of Figure~\ref{fig:k2composite}, this shelf in the K2 distribution is smoothed over by intermediate-age $\alpha$-poor stars, which we see (moving down the columns of Figure~\ref{fig:agevz}) are dominant at $|Z|>0.5$~kpc. Identically, when we look at stars in the Kepler field in the range $1{\rm\,kpc} < |Z| < 2{\rm\,kpc}$, we see that the proportion of younger (versus intermediate-age) $\alpha$-poor stars has decreased, and the overall distributions in these bins more closely resemble those same bins for the K2 fields.

In \cite{LYLu+2022}, it is shown that the $\alpha$-rich population (which we know from, e.g., \citealt{Hayden2015}, among others, to be more prominent at higher $|Z|$ due to being an older population that has experienced more dynamic heating) lacks a clear age-metallicity relation. Over time, stars at a given Galactic birth radius tend to be born with increasingly lower [$\alpha$/M] and higher [Fe/H] than the initial $\alpha$-rich population born at that radius. In a gas-rich environment with efficient star formation, this happens quickly{, creating a radial trend in stellar metallicity distribution such that metallicities are higher in the center of the Galaxy than the exterior after the same amount of time.} Then, though $\alpha$-poor stellar abundances are a function of Galactic birth radius and age, radial migration (e.g., as described by \citealt{sellwoodbinney2002}, \citealt{blandhawthorn+2010}) is sufficient to mostly wipe out the present-day radial dependence{, giving rise to the $\alpha$ bi-modality} (this is also discussed by, e.g., \citealt{sharma+2021}). {An age-$|Z|$ trend in the $\alpha$-poor population would be expected from either vertical heating mechanisms or imprinted from upside-down disc formation \citep[e.g.,][and references therein]{bird+2021}.} For our data, we seem to be seeing this same phenomenon, with a lack of strong age trends with $R$, but trends in $|Z|$ reflecting the evolution of the $\alpha$-poor sequence over time. {We stress that while the age distributions of the $\alpha$-rich and $\alpha$-poor populations are typically older and younger, respectively, the $\alpha$-poor population age distribution varies with Galactic height in K2 and Kepler data. What correlations there may be in the ages of the $\alpha$-rich population with position or metallicity are not obvious with the precision afforded by either K2 or Kepler data. Ultimately, the $\alpha$ bi-modality is a multi-variate problem, which requires models that take into account metallicity- and position-dependent SFHs to begin to understand. We explore the K2 $\alpha$ bi-modality in this context in what follows.}
%Therefore, though it is true that stars with differing compositions can be found to have the same age, or vice versa, the combination of Figure~\ref{fig:agevz} with the known Galactic trends in chemistry suggests that the empirical differences in age that we see between the bulk K2 and Kepler samples are likely showing \emph{real} age differences between similar, yet distinct, stellar populations that are being probed in different regions of the Galaxy.}

\subsection{Comparison to modeled populations} \label{ssec:models}

In order to test whether the predictions of various Galaxy formation theories are consistent with our data, we have compared our results to Galactic evolution and chemical enrichment models from \cite{modelsource1}.
These Galactic chemical evolution models predict elemental abundances of stellar populations across the disk given two different models of the Galactic SFH. The inside-out SFH model has a generic "rise-fall" shape, while the late-burst model follows the same formalism, but with a slow Gaussian-shaped bump in the rate $\sim$2~Gyr ago (motivated by the observations of \citealt{isern2019} and \citealt{mor+2019}). Stellar populations are subject to radial migration and vertical heating over their lifetimes under a prescription based on the h277 hydrodynamic simulation (e.g., \citealt{christensen+2012}). We refer to \S2.2 and \S2.5 of \cite{modelsource1} for further details.

The product of these models is a table of stellar populations along with each population's age, stellar mass, birth and present-day Galactic radius and height above (or below) the Galactic plane, [Fe/H], and [O/Fe].\footnote{Oxygen, being an $\alpha$ element, behaves similarly in tracking the Galactic $\alpha$ bi-modality.} % nice description
In order to convert these data into a selection that roughly mimics our data from K2 GAP, we assigned every population of stars from the model a weight corresponding to the fraction of stars in that population that are in the mass range that would be occupied by stars on the RGB at that population's present-day age, given the initial-mass function from \cite{Salpeter1955}. The RGB mass range was calculated assuming that a star spends $\sim$10\% of its main-sequence lifetime on the RGB,\footnote{Though this may be a slight overestimate of the RGB lifetime, adjusting this estimate by $\pm5\%$ negligibly affects our results.} and given the mass-luminosity relationship of $M \propto L^{3.8}$ (for the data from \citealt{masslumref1}, fit in \citealt{pinsonneault_ryden_2023}).
% Marc: this is likely to be an overestimate of the fractional lifetime - Jamie can actually measure this with her tracks, and I'd be curious to see what we find. This would be a useful number to have in our heads! Ditto the RC lifetime, which could be drawn from something like the published MIST tracks.  I'd define "log g = 3.3" as the base of the RGB, as that is basically the lower gravity range of asteroseismic surveys. % I think that it drops off much faster with increased mass than this, e.g. tau (RGB)/tau(MS) is a strong function of M.
%In order to match the radial and vertical distributions of stars from the APO-K2 sample, we created bins 2 kpc wide from $R = 0{\rm\,kpc}$ to $R = 16{\rm\,kpc}$ and recorded the number of stars in the APO-K2 sample that fall within each of these bins, $N_{\rm bin}$, as well as the 25th and 75th percentiles of the distance away from the plane for these stars, $Z_{25\%}$ and $Z_{75\%}$.
In order to match the radial distribution of stars from the APO-K2 sample, we created bins 2 kpc wide from $R = 0{\rm\,kpc}$ to $R = 16{\rm\,kpc}$ and recorded the number of stars in the APO-K2 sample that fall within each of these bins, $N_{\rm bin}$. 
We then cut the models down to sub-samples to match the bounds of each bin in $R$, then draw $N_{\rm bin}$ populations from the sub-sample using the weights defined above.
Finally, in order to produce model data sets that roughly mimic our observed sample,  the abundances and ages of these populations have been randomly perturbed using the mean uncertainties for these parameters in our catalog.

\begin{figure*}
    \includegraphics[width=\textwidth]{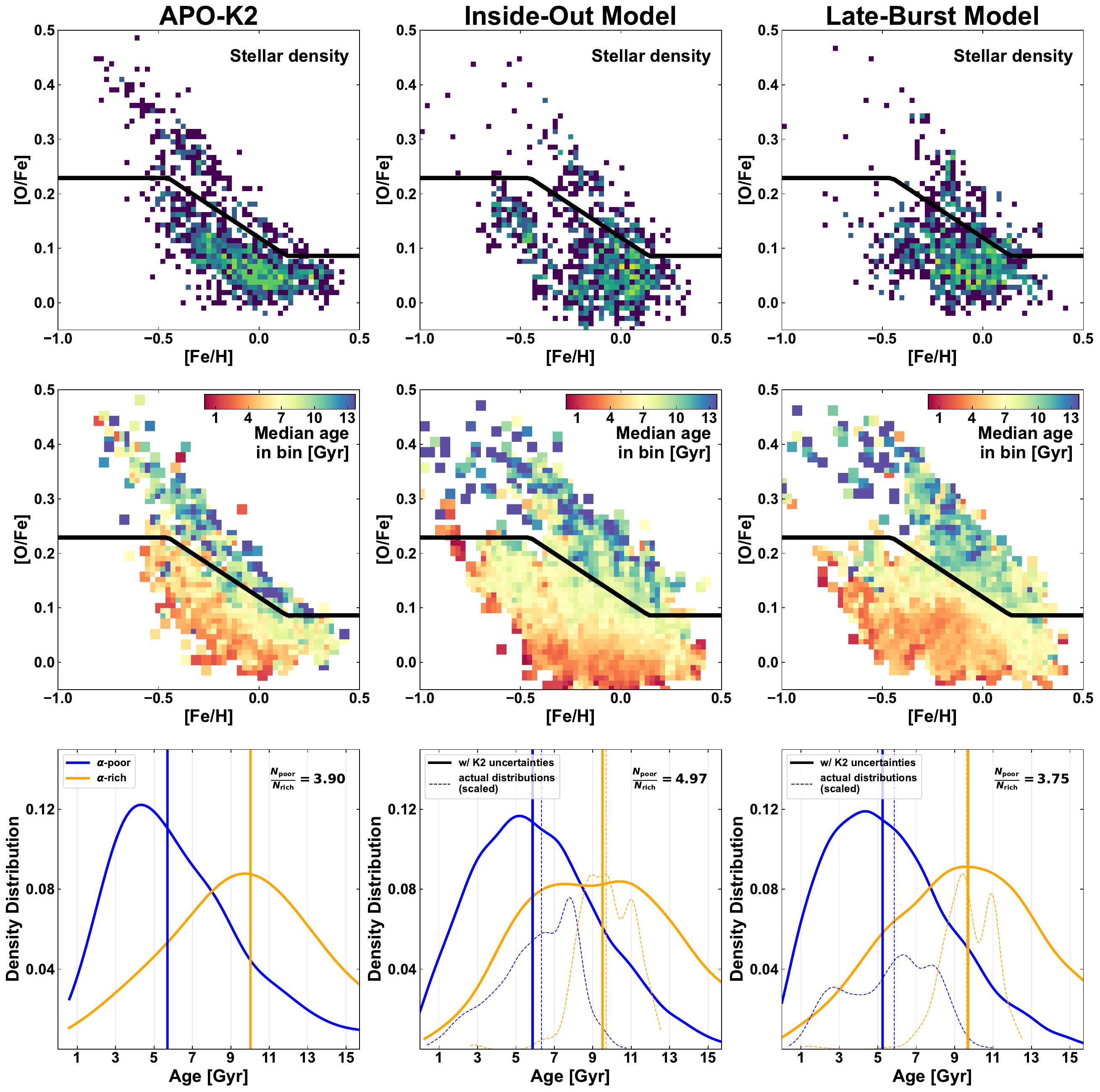}
    \caption{\textit{Top row:} Number density distributions for RGB stars in the APO-K2 Catalog within 0.5 kpc of the Galactic plane compared to two models from \protect\cite{modelsource1}: a stellar population drawn from inside-out formation models and a stellar population drawn from "late-burst" formation models in [Fe/H] vs. [O/Fe] space. The scaling is logarithmic. The black line drawn across the distributions represents the ridge-line separating the $\alpha$-rich and $\alpha$-poor populations in the \ktwo\ data. \textit{Middle row:} The median age of stars within each bin in the corresponding plot on the top row. \textit{Bottom row:} Age density distribution KDE curves for the $\alpha$-poor (blue) and the $\alpha$-rich (orange) populations from each respective sample. The area under the $\alpha$-poor KDE is normalized independently from the $\alpha$-rich. For the model populations, we plot the actual age distributions (i.e., ages from the models without APO-K2-like uncertainties) as dashed curves, which are normalized to fit under the solid curves. Vertical lines mark each chemical population's median age---this does not appear for the $\alpha$-rich actual age distributions as the median is about identical to that with uncertainties. Within each panel, we give the value for the ratio of the number of $\alpha$-poor stars to the number of $\alpha$-rich stars, $N_{\rm poor}/N_{\rm rich}$. The total number of stars drawn from each model matches the number of stars in the plotted subset of APO-K2.}
    \label{fig:model_hists}
\end{figure*}
Histograms of the densities of the resulting populations, in [Fe/H] vs. [O/Fe] space, are shown in the top row of Figure~\ref{fig:model_hists} for the APO-K2 sample (left column) and the inside-out (middle column) and late-burst (right column) star formation scenarios. Here, we have only considered stars from the APO-K2 sample that are within 0.5 kpc of the Galactic plane. %, as the models we have used assume all stars are born at $Z=0$. \jwj{I'd simply cut the sentence off at ``of the Galactic plane.'' It's less that they're born at~$z = 0$ and more that the ISM is chemical homogeneous in the vertical direction, so the birth~$\left|z\right|$ just doesn't matter, so we don't even bother tracking it. Due to star particle counts in h277, the final midplane distances are subject to sampling noise as~$\left|z\right|$ gets large, so from a model perspective, that cut is just where the~$\left|z\right|$'s are the most reliable.}

One obvious difference between the models and the data from these plots is the existence of the $\alpha$ bi-modality in this chemical space; though there are arguable over-densities of $\alpha$-rich and $\alpha$-poor stars in these plots, the distinct ridge between these populations is less pronounced than it is observationally.
The failure to reproduce the separation between the two sequences is a known problem with the \cite{modelsource1} models, whereby more intermediate [O/Fe] stars are predicted than are observed (see \citealt{modelsource1}, Figure~12). Despite this shortcoming, however, the models do accurately predict the observed variations in the abundance distributions between different Galactic regions.
In addition to the ridge-line in the bi-modality, the $\alpha$-poor loci of these models are less constrained to a clear sequence versus the APO-K2 sample, and include additional over-densities of $\alpha$-poor stars with ${\rm [Fe/H]} \lesssim -0.5$ and $\alpha$-rich stars with ${\rm [Fe/H]} \gtrsim 0$. %What really is most notable, though, is that---at least for our scheme of drawing populations from these samples---the proportion of $\alpha$-poor stars to $\alpha$-rich stars is much higher (about 3.9 in both cases) versus what we see across the \ktwo\ campaigns, where there is an approximately equal number of stars in both populations.
%\jwj{Suggestion to fit in somewhere with this paragraph: The failure to reproduce the separation between the two sequences is a known problem with the~\citet{modelsource1} models, whereby more intermediate [O/Fe] stars than observed are predicted (see their Fig. 12). Despite this shortcoming, however, the model does accurately predict the variations in the abundance distributions between different Galactic regions.}

Regardless of these differences, both the inside-out and late-burst models seem to do qualitatively well at matching the age distribution of stars within this chemical space (Figure~\ref{fig:model_hists}, middle row), including an old $\alpha$-rich population with a scattered age distribution and an $\alpha$-poor population that transitions from intermediate to young ages as a function of decreasing $\alpha$ abundance. Translated to 1D histograms for the age distributions for the $\alpha$-rich and -poor populations (bottom row), we also see quantitative agreement. Considering the observational uncertainties associated with APO-K2, the median ages of these model populations are in excellent agreement with our data, and---though models do predict different proportions of stars in each of the respective chemical populations---the KDE curves seem to qualitatively resemble those of the data. Therefore, despite some remaining discrepancies between the data and models, we are unable to rule out the plausibility of either of these scenarios.

Comparing the actual, pre-perturbation age distributions from the models (plotted as dashed curves under the inside-out and late-burst data in the bottom row of Figure~\ref{fig:model_hists}) to the ages produced when including K2-like uncertainties (solid curves in Figure~\ref{fig:model_hists}), we see significant differences not only in the relative width of the distributions, but also in the shapes of the distributions. The age uncertainty convolution process, of course, will widen the distributions, washing out any substructure that may (or may not) be present. It will also tend to skew the distributions due to the fact that the stars have roughly constant fractional age uncertainties, not absolute age uncertainties. Nonetheless, the median age of each chemical population remains relatively unchanged after convolution, and should allow for meaningful inferences on true, population-level median ages.
%\jwj{Based on this, would it be correct to simply say that the measurement uncertainties dominate the widths of the distributions, washing out any substructure that may (or may not) be present?}

\section{Discussion and Conclusions} \label{sec:conclusions}

%{\color{red}
%\begin{itemize}
%    \item The distributions of ages in the \ktwo\ and \kepler\ fields seem somewhat consistent with the populations within these fields being representative of the same underlying population.
%
%    \item The differences in age between the \ktwo\ and \kepler\ populations seems to be attributable to the higher average metallicity of stars in the \kepler\ field versus metallicities across all of the \ktwo\ fields. \jcz{Again, I think it's just the Z-effect, where K2 is sampling a wider range in Z and there is a trend with Z and age. Can clarify the metallicity bit by looking at the scatter plot of feh v. Zmax --- see previous comments.}
%
%    \item The predictions for stellar age distributions as functions of chemical enrichment for inside-out and late-burst Galactic formation models, considering observational uncertainties, seem consistent with the age distributions for stars in the APO-K2 data set located close to the Galactic plane.
%\end{itemize}
%}

In \S\ref{ssec:k2vkep} and \S\ref{ssec:gal_pos}, we compared the distributions of ages in the \ktwo\ and \kepler\ fields, and found differences in the ages between these populations are related to differences in $|Z|$ and $R$, which {may be} related to the generally lower mean metallicity of the Kepler sample. In fact, when we re-{sample} the K2 {stars to reproduce the \feh\ distribution of the Kepler stars,} we see a significant decrease in the offsets between the median ages of the populations. {Re-sampling by $(R, |Z|)$ yields similar results.}

Considering the relationship between the mean \feh\ of a given population and $(R, |Z|)$ (as shown by, e.g., \citealt{Hayden2015}, \citealt{imig+2023}), it is reasonable to assume that these two avenues of re-weighting are achieving the same ends. This relationship between chemistry and position---and the extended relationship that we have confirmed with age---demonstrates the importance of accounting for all properties of a given stellar population before applying these ages more broadly, particularly in the case of the ages of the Kepler field stars. I.e., otherwise similar stars found in different locations of the Galaxy may have different ages, which is related specifically to gradients in the SFH and the migration of stars across the Galaxy. %I.e., stellar metallicity not only influences the measured main sequence lifetime of a single given star, but may also relate specifically to measureable gradients in the SFHs of both the $\alpha$-poor thick disc and $\alpha$-rich thin-disc across the Galaxy.

%\jwj{I suggest expanding the first sentence here into its own paragraph, since this is the main comparison of section~\ref{ssec:models}. The models reproducing the alpha bi-modality is worth its own discussion, but it's sufficiently separate from what you've done. The main difference between the IO and LB models is the recent SFH, so what we really learn here is that the age uncertainties in your models are too large to constrain the Galactic SFH. Suggestion: In~\S~\ref{ssec:models}, we compared the age distributions of APO-K2 stars to~\citeauthor{modelsource1}'s~\citeyearpar{modelsource1} Galactic chemical evolution models. Their late-burst SFH differs from their inside-out SFH only in that it includes a recent, slow burst of star formation motivated by the observations of Isern (2019) and Mor et al. (2019). The predictions of both models are broadly consistent with our APO-K2 sample due to the substantial age uncertainties. We therefore cannot distinguish between these models of the recent SFH in the Milky Way.}
%In \S\ref{ssec:models}, we show that the predictions the distributions of stellar ages as functions of chemical enrichment from inside-out and late-burst Galactic formation models seem consistent with (or, at least, cannot be ruled out by) the age distributions for stars in the APO-K2 data set. 
In \S\ref{ssec:models}, we compared the age distributions of APO-K2 stars to \citeauthor{modelsource1}'s \citeyearpar{modelsource1} Galactic chemical evolution models. Their late-burst SFH differs from their inside-out SFH only in that it includes a recent, slow burst of star formation (observationally motivated by \citealt{isern2019} and \citealt{mor+2019}).
Broadly, the predictions of both models are consistent with our APO-K2 sample. However, due to the substantial age uncertainties for our data, we are unable to distinguish between these models in terms of the Galaxy's recent SFH, and, therefore, neither model can be preferred nor ruled out. 

%\jwj{I suggest making this its own paragraph and perhaps editing it down since the {\it origin} of the alpha bimodality isn't central to this paper.}
However, it is also important to note that these inside-out and late-burst models from \cite{modelsource1} fail to fully recover the MW's $\alpha$ bi-modality. Much has been said in the recent decades for possible histories able to produce both the kinematic and the chemical thin- and thick-disks.
\cite{Clarke2019} shows that bursty episodes of brief, high star formation in simulated galaxies are able to account for both the chemical bi-modality and the associated age bi-modality of the MW. This finding seems especially promising, considering that these clumpy star-forming regions are also observed in disk galaxies at high redshift (e.g., as first identified by \citealt{CowieHuSongaila1995}, \citealt{vandenBergh1996}). {However, the fine-grained age predictions from this model do not agree with the data (see \citealt{Clarke2019}, Figure 12). On the other hand, t}wo- and three-infall models---where the bi-modality is primarily produced by the separation of two epochs of star formation driven by distinct infall events of pristine gas into the Galaxy---have been shown capable of recovering the general age-$\alpha$ relation \citep{Chiappini2015,Spitoni2019} and age-$Z_{\rm max}$ relation \citep{Spitoni2022}, at least as they are observed in the \kepler\ field/the Solar circle (using data from, e.g., \citealt{SilvaAguirre+2018}, \citealt{Ting&Rix2019}, \citealt{Leung+Bovy2019}).
%\jcz{I think this following sentence should be replaced with one that says something like Tests of the predicted age distributions as a function of chemical space with fine-grained age binning as presented here would place interesting constraints on this class of models.} 
{As it stands, in-depth comparisons of the predictions of these models to those of parallel evolution models (such as those of \citealt{modelsource1}) have not been done. These comparisons, done via tests of predicted age distributions as a function of chemical space with fine-grained age binning (as presented here) would place interesting constraints on the relative strengths of each class of models versus observed trends.}
%{As it stands, many of the specific predictions of these models are too particular or difficult to confirm with available age data. Tests of predicted age distributions as a function of chemical space, with the limits of available observational data in mind, will be necessary to place interesting constraints on this class of models.} \jtw{needs to be some comparison to other classes of models}
%{As it stands, many of the specific predictions of these models are too ambiguous or difficult to confirm with available age data. Tests of predicted age distributions as a function of chemical space, with fine-grained age binning as presented here, would place interesting constraints on this class of models.}
%However, many of the specific predictions of these models are either too ambiguous to be tested with available age data or make predictions for the evolution of stellar populations through ${\rm [\alpha/Fe]}$-\feh\ space that have already been directly contradicted (e.g., as noted by \citealt{sharma+2021}).

In this paper, we have presented precise, asteroseismically-derived ages for a large and comprehensive sample of stars in the K2 and Kepler fields.
We have shown that, despite being rather ubiquitous across the Galaxy, the $\alpha$-rich and $\alpha$-poor populations show small variations in their ages that are at least partial functions of their locations in $(R, |Z|, {\rm [Fe/H]})$ phase-space. In-line with this, we conclude that nearly all of the differences that we find between the ages of $\alpha$-rich populations of the Kepler field versus the K2 fields are due to the larger uncertainties of the K2 data, and offsets in the age distribution of the $\alpha$-poor population are attributable to the differences in stellar Galactic position and metallicity between the samples. Although K2 affords a wider selection of stellar populations than Kepler, comparisons of the K2 ages against models as a function of ${\rm [\alpha/Fe]}$-\feh\ space are not precise enough to distinguish between scenarios with versus without recent star formation. Combining Gaia distance information with TESS asteroseismology is promising for achieving more precise ages across a wide range of Galactic environment (e.g., \citealt{aguirre+2020TESS}, \citealt{stello+2022}), which will allow for further investigations of Galactic chemical evolution models beyond the Kepler field.

\begin{acknowledgments}
Funding for the Sloan Digital Sky Survey IV has been provided by the Alfred P. Sloan Foundation, the U.S. Department of Energy Office of Science, and the Participating Institutions. 

SDSS-IV acknowledges support and resources from the Center for High Performance Computing  at the University of Utah. The SDSS website is www.sdss.org.

SDSS-IV is managed by the Astrophysical Research Consortium for the Participating Institutions of the SDSS Collaboration including the Brazilian Participation Group, the Carnegie Institution for Science, Carnegie Mellon University, Center for Astrophysics | Harvard \& Smithsonian, the Chilean Participation Group, the French Participation Group, Instituto de Astrof\'isica de Canarias, The Johns Hopkins University, Kavli Institute for the Physics and Mathematics of the Universe (IPMU) / University of Tokyo, the Korean Participation Group, Lawrence Berkeley National Laboratory, Leibniz Institut f\"ur Astrophysik Potsdam (AIP),  Max-Planck-Institut f\"ur Astronomie (MPIA Heidelberg), Max-Planck-Institut f\"ur Astrophysik (MPA Garching), Max-Planck-Institut f\"ur Extraterrestrische Physik (MPE), National Astronomical Observatories of China, New Mexico State University, New York University, University of Notre Dame, Observat\'ario Nacional / MCTI, The Ohio State University, Pennsylvania State University, Shanghai Astronomical Observatory, United Kingdom Participation Group, Universidad Nacional Aut\'onoma de M\'exico, University of Arizona, University of Colorado Boulder, University of Oxford, University of Portsmouth, University of Utah, University of Virginia, University of Washington, University of Wisconsin, Vanderbilt University, and Yale University.

S.M.\ acknowledges support by the Spanish Ministry of Science and Innovation with the Ramon y Cajal fellowship number RYC-2015-17697 and the grant number PID2019-107187GB-I00, and through AEI under the Severo Ochoa Centres of Excellence Programme 2020--2023 (CEX2019-000920-S).
D.S. is supported by the Australian Research Council (DP190100666).
R.A.G acknowledges the support from the PLATO Centre National D'{\'{E}}tudes Spatiales grant.
\end{acknowledgments}


\vspace{5mm}
%\facilities{Kepler}


%\software{astropy \citep{2013A&A...558A..33A,2018AJ....156..123A},  
%          Cloudy \citep{2013RMxAA..49..137F}, 
%          Source Extractor \citep{1996A&AS..117..393B}
%          }


%% Figure~and Table counter will not reset.

\appendix

\section{Asteroseismic Data and Catalog Pipeline} \label{app:data}
Throughout this paper, we have described our use and modifications of various published data sets, namely the APO-K2 Catalog as presented by \akccite, which itself is constructed primarily using the K2 GAP DR3 catalog \citep{zinn+2022} and APOGEE DR17 \citep{SDSSdr17}. From their origins as K2 imaging to the ages we present in this paper, these data have come together through the collective efforts of many different individuals and teams. In this appendix, we summarize the process used to build and calibrate the catalogs presented here and by \akccite.

\subsection{Data processing} \label{app:data:start}
%Getting parameters from lightcurves, combining pipelines \\
The K2 Galactic Archaeology Program selected red giant asteroseismic targets for K2 observations based on simple color-magnitude selection criteria, with typical K2 campaign selection criteria being (J-$\mathrm{K_s}$) > 0.5 and (9 < V < 15). Additional targets were selected based on previous spectroscopic identification from surveys, including APOGEE. Ultimately, more than 110,000 targets were observed from these target lists in campaigns C1-C8 and C10-C18. The selection process is discussed in more detail by \cite{sharma+2022}.

Light curves were generated using EVEREST \citep{luger+2018}, which uses pixel-level data to remove systematics associated with K2's degraded pointing compared to that of Kepler. The K2 GAP light curves were reduced and calibrated in a manner appropriate for asteroseismology, including high-pass filtering and inpainting missing data \citep{pires+2015}. Members of the Kepler Asteroseismic Science Consortium (KASC)\footnote{\url{https://kasoc.phys.au.dk}} analyzed each star, resulting in a set of asteroseismic parameters from up to six independent pipelines. The final adopted values in the K2 GAP DR3 catalog are averages of the frequency of maximum oscillation power ($\nu_{\mathrm{max}}$) and the large frequency separation ($\Delta\nu$) values across the pipelines (and campaigns, if a star was observed multiple times), with discrepant pipeline results rejected with sigma clipping \citep{zinn+2022}.

%The APO-K2 catalog combines the K2 GAP asteroseismic catalog with spectroscopic data from APOGEE during the fourth phase of the Sloan Digital Sky Survey (SDSS-IV) \citep{Blanton2017} and analyzed in its seventeenth (and final) data release (DR17) \citep{Accetta2021}. Compared to DR16 \citep{jonsson2020}, DR17 contains 675,000 APOGEE targets over an additional two years of multi-hemisphere SDSS observations.

\subsection{Crossmatches} \label{app:data:cross}
To build our catalog, we started with Table 6 from \cite{zinn+2022}, which has a row for each star identified by its Ecliptic Plane Input Catalog (EPIC) ID and the K2 campaign it was observed during, providing the asteroseismic \numax\ and \dnu\ values from each of the six pipelines. Each unique EPIC ID was matched to 2MASS IDs \citep{Skrutskie2006} using the \texttt{kepler.k2\_epic} table in the MAST Query/\texttt{CasJobs}\footnote{\url{https://mastweb.stsci.edu/mcasjobs/}} database, which gives a 2MASS match for every target with an EPIC ID. These 2MASS IDs were then used to match to the APOGEE DR17 catalog, where stars are given APOGEE IDs that are equivalent to 2MASS IDs. As-is, the APOGEE catalog will have more than one entry for some 2MASS IDs due to the same source being targeted in distinct observing programs. In order to assure a one-to-one crossmatch, we first sorted the APOGEE table by \texttt{SNR}, then dropped the duplicate 2MASS entries with the lower \texttt{SNR} value. From this crossmatched table, we removed all objects for which there was no measured $\nu_{\rm max}$, $\Delta\nu$, or [Fe/H].

\subsection{$\Delta\nu$ and $\nu_{\rm max}$ Calibration} \label{app:data:cal1}
% Sharma, Gaia
Asteroseismic masses and radii are calculated according to scaling relations following the notation of \cite{sharma+2016}:
\begin{equation}
    \frac{M}{M_\odot} \approx \left( \frac{\nu_{\rm max}}{f_{\nu_{\rm max}}\nu_{\rm max, \odot}} \right)^3 \left( \frac{\Delta\nu}{f_{\Delta\nu}\Delta\nu_{\odot}} \right)^{-4} \left( \frac{T_{\rm eff}}{T_{\rm eff, \odot}} \right)^{3/2} \equiv \kappa_M \left( \frac{T_{\rm eff}}{T_{\rm eff, \odot}} \right)^{3/2},
\end{equation}
\begin{equation}
    \frac{R}{R_\odot} \approx \left( \frac{\nu_{\rm max}}{f_{\nu_{\rm max}}\nu_{\rm max, \odot}} \right) \left( \frac{\Delta\nu}{f_{\Delta\nu}\Delta\nu_{\odot}} \right)^{-2} \left( \frac{T_{\rm eff}}{T_{\rm eff, \odot}} \right)^{1/2} \equiv \kappa_R \left( \frac{T_{\rm eff}}{T_{\rm eff, \odot}} \right)^{1/2}.
\end{equation}
These scaling relations are approximations linked to solar values and require calibration factors, $f_{\Delta\nu}$ and $f_{\nu_{\mathrm{max}}}$, because \dnu\ does not scale exactly as stellar mean density \citep{White2011} and \numax\ does not scale exactly as g/$T_{\rm eff}$ \citep{brown1991,kjeldsenbedding1995}.
These calibration factors vary by star and depend on, in part, the star's temperature and metallicity.

For APO-K2, we used APOGEE DR17 temperatures and metallicities to update $f_{\Delta\nu}$ using \texttt{Asfgrid}\footnote{\texttt{Asfgrid} is publicly available at \url{http://www.physics.usyd.edu.au/k2gap/Asfgrid/}} \citep{sharma+2016} with the low-mass, low-metallicity extension \citep{Stello2022}. Specifically, we used \texttt{Asfgrid} to perform a multi-dimensional interpolation of $f_{\Delta\nu}$ using evolutionary state (see below), a Salaris-corrected metallicity using APOGEE [$\alpha$/M] \citep{salaris+1993}, APOGEE temperature, \dnu, and \numax.

The K2 GAP DR3 asteroseismic data that feed into the APO-K2 catalog were additionally calibrated using Gaia parallaxes inferred from bulk stellar motions from Gaia DR2 \citep{gaia2} according to the methodology detailed in \cite{schonrichmcmillan+eyer2019} and accounting for selection functions according to \cite{schonrich+aumer2017}. This technique corrects for parallax bias, and indicates $\sim 10$~$\mu$as positional variations in the Gaia parallax zero-point across K2 campaigns.
The resulting Gaia parallaxes were then compared to asteroseismic parallaxes, which can be computed from asteroseismic radii in combination with the Stefan-Boltzmann law. $f_{\nu_{\mathrm{max}}}$ was defined to bring the asteroseismic parallaxes into agreement with the Gaia parallaxes. A separate value was computed for RGB and RC stars, in recognition of their different stellar structure (and therefore potentially different asteroseismic systematics) as well as their potentially different selection functions (and therefore potentially different Gaia parallax systematics).

\subsection{Evolutionary States} \label{app:data:estates}
With a long enough time baseline, the evolutionary state of stars can be determined directly through an asteroseismic analysis of light curves (e.g., \citealt{bedding+2011}). Though this is possible for the original Kepler/APOKASC data (e.g.; \citetalias{APOKASC2}; \citealt{apokascstates}; M. Pinsonneault et al. 2024, in preparation 2024), and some work has been done on doing the same light curve analysis for shorter-baseline data using neural networks (e.g., \citealt{Hon+2018}), these methods are still far from ideal for the relatively noisy K2 data.

\begin{figure}
    \includegraphics[width=\columnwidth]{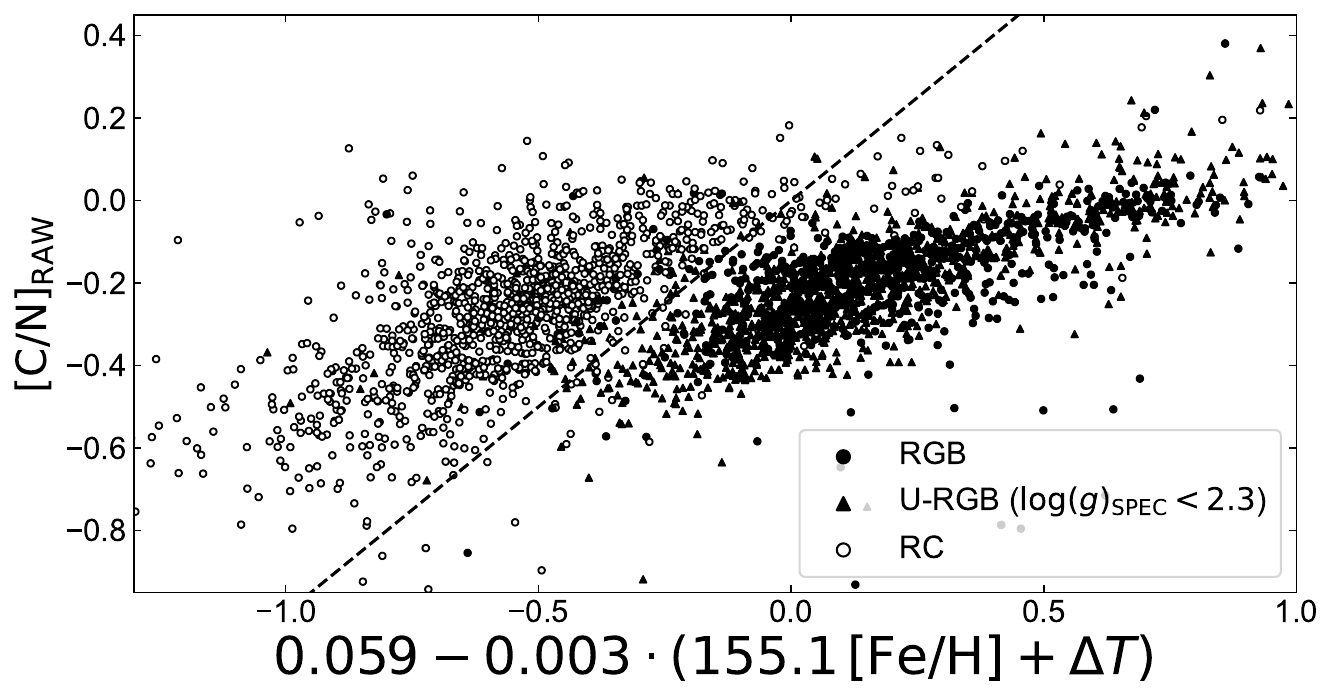}
    \caption{A random sample of asteroseismically-classified RGB (filled) and RC (unfilled) stars from the APOKASC-3 catalog, plotted in the space of the spectroscopic fit. The $y$-axis is the uncalibrated APOGEE [C/N] and the $x$-axis is an empirical surface gravity- and metallicity-dependent parameter (see text for details). Stars with low surface gravities and therefore assumed to be RGB/AGB stars spectroscopically are also shown. The dashed line is one-to-one, under which stars would be spectroscopically classified as RGB and above RC.}
    \label{fig:ak3_states}
\end{figure}
An alternative is to determine the evolutionary states of stars spectroscopically using cuts in temperature, surface gravity, and element abundances (specifically, [Fe/H] and [C/N]; e.g., \citealt{Bovy2014}, \citealt{apokascstates}). To do this, we first define a "reference" temperature, which is defined as the typical effective temperature we would expect for an RGB of a given metallicity and surface gravity:\footnote{The uncalibrated metallicity ([Fe/H]$_{\rm RAW}$) and surface gravity ($\log(g)_{\rm SPEC}$) from APOGEE DR17 are used as the calibrated versions are dependent on APOGEE's own evolutionary state determinations.}
\begin{equation}
    T_{\rm ref} = \alpha + \beta \, {\rm [Fe/H]_{\rm RAW}} + \gamma \, (\log(g)_{\rm SPEC} - 2.5).
\end{equation}
Spectroscopic parameters are taken from APOGEE DR17, and $\alpha$, $\beta$, and $\gamma$ are fit parameters that are determined through a nonlinear least-squares fit for stars classified as RGB in the APOKASC-3 catalog. After determining these values ($\alpha = 4427.1779$, $\beta = -399.5105$, $\gamma = 553.1705$), a Monte Carlo optimization is used to find values for $a$, $b$, and $c$ in a second expression:
\begin{equation} \label{eq:stateexp}
    a - b \, (c \, {\rm [Fe/H]_{\rm RAW}} + T_{\rm eff}^{\rm SPEC} - T_{\rm ref}) - {\rm [C/N]_{\rm RAW}},
\end{equation}
for which at least 98\% of stars with a value less than 0 have an asteroseismic RGB classification (Figure \ref{fig:ak3_states}). We found best-fit values of $a = 0.05915$, $b=0.003455$, and $c=155.1$. This expression is then applied to our APO-K2 data set, assigning all stars for which \ref{eq:stateexp} $<0$ as RGB and $>0$ as RC. The exception is for stars on the upper-RGB (U-RGB), which we have conservatively defined as $\log(g) < 2.3$. These stars fall in a region of a Kiel diagram beyond what is possible for the RC, and so have all been classified as RGB, regardless of their value for \ref{eq:stateexp}.

\subsection{Selection Function} \label{app:data:selfun}
% They layers of the selection function 
Due to being the product of a cross-match between K2 GAP and APOGEE DR17, the APO-K2 catalog is subject to selection and targeting choices made by both of these surveys. By understanding the selection function of the underlying samples, \akccite\ assessed the completeness of the sample and existing effects on the distributions of stellar parameters by comparing the selection function for the whole APO-K2 sample to that of K2 GAP. This selection function is defined in bins of mass and stellar metallicity, giving the ratio of the number of stars in the APO-K2 sample versus the K2 GAP sample for each bin. These ratios can then be used as weights when analyzing the APO-K2 sample.

This same comparison is also done between the APO-K2 sample and a set of simulated stars drawn from a parent mock Galactic asteroseismic red giant population generated with \texttt{Galaxia} \citep{sharma+2011}. These simulated stars are drawn in color and magnitude space in accordance with the original K2 GAP targeting selection function, and then have an expected asteroseismic selection function applied in order to only retain stars with a $>90\%$ probability of having detectable asteroseismic signals \citep{sharma+2022}.

\subsection{Ages} \label{app:data:ages}
Our ages in this paper use the methodology and code that some of these authors first presented in \citetalias{warfield+2021}. Each star is defined by its values for $\log({\rm Mass})$ (from asteroseismology), [Fe/H] and [$\alpha$/M] (from APOGEE DR17), and the 1$\sigma$ uncertainties for each of these, which are then used as look-up parameters in a regular four-dimensional grid---including $\log({\rm age})$---constructed from sets of stellar evolutionary tracks.

Our stellar evolutionary tracks were generated originally by/for \cite{Tayar2017} using the \texttt{Yale Rotating Evolution Code} \citep{yrec1, yrec2}. Given tabulated values for stellar mass,\footnote{These tracks do not apply any mass loss, and so treat the birth mass as the present-day mass.} present-day surface element abundances, surface gravity, and temperature, these tracks provide the age of a star when leaving the main sequence/joining the RGB (the MSLT). From these tracks, we created three sets of grids at fixed $\log(g)$ values of 3.30, 2.50, and 1.74 (chosen as discrete values in the tracks that approximately separate and bracket the low-luminosity giants from the luminous giants), and assuming a solar He abundance of 0.272683 and solar mixing length of 1.72. These grids are tabulated at mass values between 0.6 M$_\odot$ and 2.6 M$_\odot$, with 0.1 M$_\odot$ steps; [Fe/H] values between -2.0 and 0.6, with 0.2 dex steps; and [$\alpha$/M] values of 0.0, 0.2, and 0.4. 

We chose this approach because the MSLT is well-defined as a function of mass. The age of a star at a specific location on the RGB requires at least accounting for $T_{\rm eff}$, which is strongly correlated to age in this regime. However, this also leads to small uncertainties in temperature (or systematic offsets between model and data zero-points) being associated with large swings in age, with a 50~K uncertainty in $T_{\rm eff}$ (typical of the APOGEE data) corresponding to an $\sim70\%$ uncertainty in age (\citealt{Tayar2017}, \citetalias{warfield+2021}). For these reasons, and considering also that the amount of time spent on the RGB ($\sim10\%$ of the MSLT) is smaller than our stellar age uncertainties (even for the APOKASC sample), the MSLT, as determined via mass, serves as a very consistent age measurement across both asteroseismic samples.

Given values of $(x,y,z) \equiv (\log{M}, {\rm [Fe/H]}, {\rm [\alpha/M]})$, we can then find an associated value for $w \equiv \log({\rm age})$ from our model grid via four-point, four-dimensional Lagrange interpolation. In two dimensions, four-point Lagrange interpolation works by constructing a Lagrange interpolation polynomial of the form:
\begin{equation}
    L(x) = \sum_{j=0}^3 w_j l_j(x),
\end{equation}
where
\begin{equation}
    l_j(x) = \prod_{0 \leq m \leq 3 \atop m \neq j} \frac{x - x_m}{x_j - x_m}
\end{equation}
are the Lagrange basis polynomials. For a set of four ordered pairs $(x_0, w_0), ..., (x_3, w_3)$, this equation gives an estimate for values of $w = L(x)$ for any given $x \in (x_0, x_3)$ while keeping that $L(x_j) = w_j$. Expanded to four-dimensions, this then becomes:
\begin{equation} \label{eq:4pt}
    L(x,y,z) = \sum_{j=0}^3 \sum_{k=0}^3 \sum_{m=0}^2 w_{j,k,m} l_j(x) l_k(y) l_m(z).
\end{equation}
Using a Monte Carlo method to produce 5,000 $(x,y,z)$ tuples for each star (assuming a Gaussian error distribution for each parameter), we then used Equation \ref{eq:4pt} to make two $\log({\rm age})$ estimates for each tuple, one with each of the tracks for the $\log{g}$ values that bracket the input star's $\log{g}$. For this calculation, four nearest-neighbor points were chosen from the model grid so that, for each parameter, $x_0 < x_1 < x < x_2 < x_3$. The exception is for $z \equiv {\rm [\alpha/M]}$ where, since the tracks from \cite{Tayar2017} only sample three values of [$\alpha$/M], $(z_0,z_1,z_3) = (0.0, 0.2, 0.4)$ always, and so the associated basis polynomial, $l_m(z)$, is always constructed using these points.
%\jcz{perhaps say that is why $l_m$ only has 2 degrees of freedom and associate each l with a lookup variable. i would also explain why you took this approach since the main sequence age is well-defined and there is not any weird weighting of the ages due to a grid-based approach where the temperature might through the age off --- you're only dealing with mass.}

After obtaining 5,000 $\log({\rm age})$ estimates, the median and $\pm1\sigma$ were calculated in $\log$ space for each grid, and then converted to linear space for tabulation. The mean of the low- and high-$\log{g}$ grid values is used, with the difference between them added as a systematic error. However, in all cases, there is no offset between these two values.

\section{Ages on the Red Clump} \label{app:rc}
In \S\ref{ssec:cuts}, we explain our reasoning for only considering RGB stars as being due to the mass loss in the RC not being fully understood, leading to biases in the ages of these stars \citep[e.g.,][]{casagrande+2016}. This issue could theoretically be resolved with a prescription for this mass loss. However, it is not clear how---if at all---mass loss on the RGB is related to stellar age or abundance \citep{an+2019}. Absent a physics-driven understanding of mass loss that is applicable to stars on an individual basis, it may instead be possible to characterize this bias empirically for a given population. For instance, we might assume that the mean mass (or age) of $\alpha$-rich RC stars in the APO-K2 sample should be the same as that of the $\alpha$-rich RGB stars, and therefore apply the multiplicative offset between the means to each star's mass (or age) ex post facto. However, this would grant these stars very limited utility: there would be little reason to trust the "corrected" age of individual stars, and it would be difficult to separate genuine features of the age profile from a mis-accounting of mass loss. Regardless, this may still be an intriguing avenue for generally characterizing the expected magnitude of this change. 

\begin{figure}
    \includegraphics[width=\columnwidth]{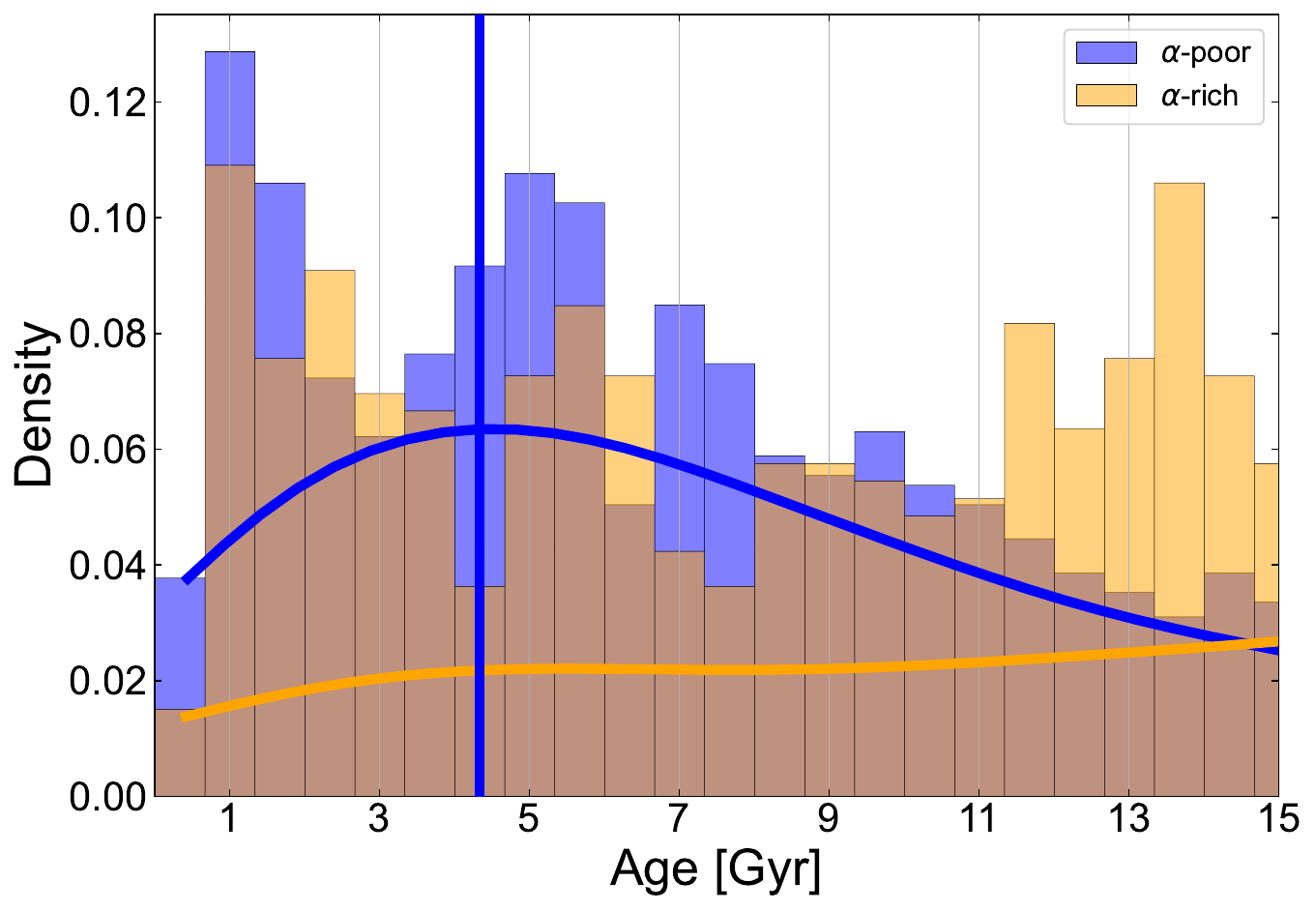}
    \caption{Age distributions for the 2298 $\alpha$-poor and 1357 $\alpha$-rich RC stars in the APO-K2 sample. These ages have been estimated using present day asteroseismic masses, uncorrected for potential mass loss.}
    \label{fig:rcnocorr}
\end{figure}
\begin{splitdeluxetable*}{cccrrrrrrrBcrrrrrrrr}

\tablecaption{The partial data table for the APO-K2 RC sample, including our ages. The complete table is available in CSV format in the online journal. The main identifier for each star is its ID in the Ecliptic Plane Input Catalog (EPIC). In addition to the columns shown here, our table contains Galactic longitude and latitudes; Galactocentric $R$ and $Z$ positions; the uncalibrated values for APOGEE $T_{\rm eff}$, $\log{g}$, [Fe/H], [$\alpha$/M], and [O/Fe], as well as the associated errors for all relevant columns. The {\tt $\alpha$-rich Flag} column has a value of {\tt 1} for $\alpha$-rich stars, {\tt 0} for $\alpha$-poor stars, and {\tt -1} for unclassified stars. \label{tab:k2_rc_table}}
\tabletypesize{\footnotesize}

\tablehead{
\colhead{EPIC ID} & 
\colhead{APOGEE ID} & 
\colhead{Gaia EDR3 Source ID} & 
\colhead{RA} & 
\colhead{DEC} & 
\colhead{$T_{\rm eff}$} & 
\colhead{$\log(g)_{\rm APO}$} & 
\colhead{[Fe/H]} & 
\colhead{[$\alpha$/M]} & 
\colhead{[O/Fe]} & 
\colhead{$\alpha$-rich Flag} & 
\colhead{Mass} & 
\colhead{Radius} & 
\colhead{$\log(g)_{\rm seis}$} & 
\colhead{$\nu_{\rm max}$} & 
\colhead{$\Delta\nu$} & 
\colhead{Age} & 
\colhead{Modal Age} \\
\colhead{} & 
\colhead{} & 
\colhead{} & 
\colhead{(Deg.)} & 
\colhead{(Deg.)} & 
\colhead{(K)} & 
\colhead{($\log({\rm cm/s}^2)$)} & 
\colhead{(dex)} & 
\colhead{(dex)} & 
\colhead{(dex)} & 
\colhead{} & 
\colhead{(M$_\odot$)} & 
\colhead{(R$_\odot$)} & 
\colhead{($\log({\rm cm/s}^2)$)} & 
\colhead{($\mu$Hz)} & 
\colhead{($\mu$Hz)} & 
\colhead{(Gyr)} & 
\colhead{(Gyr)}
}

\startdata
212532733 & 2M13122093$-$1202115 & 3621787595338351872 & 198.0872 & -12.0366 & 4644 & 2.25 & -0.296 & 0.220 & 0.307 & 1 & 0.77 & 11.33 & 2.21 & 20.4 & 3.10 & 28.2 & 27.5 \\
211163406 & 2M03582817$+$2541126 & 67156795835480448 & 59.6174 & 25.6869 & 4616 & 2.45 & -0.060 & 0.037 & 0.065 & 0 & 1.22 & 10.70 & 2.46 & 36.6 & 4.26 & 5.7 & 5.1 \\
212451149 & 2M13180826$-$1346410 & 3609158841003939968 & 199.5344 & -13.7781 & 4925 & 2.55 & -0.239 & 0.043 & 0.060 & 0 & 1.15 & 9.83 & 2.51 & 39.5 & 4.69 & 5.9 & 5.4 \\
210618287 & 2M04235747$+$1713227 & 3313965708686969728 & 65.9895 & 17.2230 & 4438 & 2.21 & -0.059 & 0.073 & 0.108 & -1 & 1.10 & 12.27 & 2.30 & 25.7 & 3.30 & 8.4 & 7.9 \\
212750556 & 2M13191487$-$0657577 & 3634414902267654400 & 199.8120 & -6.9660 & 4264 & 1.93 & -0.059 & 0.107 & 0.147 & 1 & 1.24 & 19.06 & 1.97 & 12.2 & 1.81 & 5.5 & 3.7 \\
204987895 & 2M16111595$-$1956386 & 6245469248293963648 & 242.8165 & -19.9441 & 4556 & 2.12 & -0.411 & 0.154 & 0.199 & 1 & 0.89 & 12.54 & 2.19 & 19.7 & 2.88 & 13.4 & 10.2 \\
205170082 & 2M16384287$-$1903121 & 4131201704833326592 & 249.6786 & -19.0534 & 4817 & 2.52 & -0.018 & 0.020 & 0.025 & 0 & 1.13 & 9.85 & 2.51 & 39.4 & 4.65 & 7.3 & 7.2 \\
220637427 & 2M00504808$+$0952289 & 2582124996801990528 & 12.7003 & 9.8747 & 4618 & 2.52 & -0.304 & 0.235 & 0.275 & 1 & 0.99 & 9.31 & 2.49 & 39.1 & 4.72 & 11.0 & 11.1 \\
246144695 & 2M23021866$-$0610311 & 2634924193707214080 & 345.5778 & -6.1753 & 4586 & 2.08 & -0.383 & 0.109 & 0.134 & 0 & 0.96 & 13.59 & 2.15 & 18.0 & 2.65 & 10.3 & 10.5 \\
... & ... & ... & ... & ... & ... & ... & ... & ... & ... & ... & ... & ... & ... & ... & ... & ... & ... \\
\enddata

\end{splitdeluxetable*}
In Figure~\ref{fig:rcnocorr}, we have plotted the age distributions for the $\alpha$-poor and $\alpha$-rich populations of RC stars in the APO-K2 catalog, calculated without any correction for mass loss. In one sense, these ages may still be useful. We see a slight peak in $\alpha$-rich ages at $\sim$14~Gyrs, and a gradual peak of younger $\alpha$-poor stars. However, the main difference between these ages and those for the RGB stars is that there are many more $\alpha$-rich RC stars at intermediate and young ages. Still, it is possible that this is telling us something real about these populations. For instance, we would expect more RC stars to go through mergers, and one possible explanation for the young $\alpha$-rich population (that is also observed among the RGB stars) is that these are merger products that have larger present-day masses than they had at birth, and so they are interpreted as being younger than they actually are \citep{Chiappini2015, Martig2015}. These ages can be found in Table~\ref{tab:k2_rc_table}, though we strongly encourage readers to pay heed to the limitations we have highlighted above and to exercise caution when using these data. %Stars, if classified as RC, do end up with slightly higher masses (and therefore, lower ages) than if they were classified as RGB, as a much smaller correction to \dnu\ is applied to stars on the RC compared to the RGB. \jcz{this last point should not be relevant because radii for RC corrected with a fdnu correction have good radii and masses, though there could be a 1\% systematic error w.r.t. RGB stars (pinsonneault+ 2018) discuss this. I think this should be made clear: the mass loss correction is where the problems may be, not with the fdnu. Maybe language that starts with: Note that, although red clump stars will have lower masses than they started with, the fdnu corrections, which tend to lower masses, are less important for the RC, and so their fdnu-corrected masses are larger than that of a RGB star of the same mass. something like that...}

\section{Ages for Metal Poor Stars} \label{app:mp}
Thus far, we have excluded metal poor stars from our analysis due to significant uncertainty in the calibration of the asteroseismic scaling relations at ${\rm [Fe/H]} \lesssim -1$, which, in turn, leads to systematic overestimates in the masses of these stars. \cite{Epstein2014} notably presented this issue using APOKASC-1 data, showing an $\sim$10\% overestimate for the asteroseismic masses of halo stars as compared to model predictions. Subsequent studies have shown mixed levels of systematics (see \akccite\ and references therein).

\begin{splitdeluxetable*}{cccrrrrrrrBcrrrrrrrr}

\tablecaption{The partial data table for the stars in the APO-K2 sample with ${\rm [Fe/H]} < -1$, including our ages, which have their masses calibrated and calculated using the uncalibrated APOGEE $T_{\rm eff}$ values. The complete table is available in CSV format in the online journal. The main identifier for each star is its ID in the Ecliptic Plane Input Catalog (EPIC). In addition to the columns shown here, our table contains Galactic longitude and latitudes; Galactocentric $R$ and $Z$ positions; the uncalibrated values for APOGEE [Fe/H], [$\alpha$/M], and [O/Fe], as well as the associated errors for all relevant columns. The {\tt $\alpha$-rich Flag} column is not populated for these stars, as the $\alpha$-rich and $\alpha$-poor populations are not well defined at these metallicities, though it may be assumed that these stars are more similar to the $\alpha$-rich population. \label{tab:k2_poor_table}}
\tabletypesize{\footnotesize}

\tablehead{
\colhead{EPIC ID} & 
\colhead{APOGEE ID} & 
\colhead{Gaia EDR3 Source ID} & 
\colhead{RA} & 
\colhead{DEC} & 
\colhead{$T_{\rm eff}^{\rm uncal.}$} & 
\colhead{$\log(g)_{\rm APO}^{\rm uncal.}$} & 
\colhead{[Fe/H]} & 
\colhead{[$\alpha$/M]} & 
\colhead{[O/Fe]} & 
\colhead{Mass} & 
\colhead{Radius} & 
\colhead{$\log(g)_{\rm seis}$} & 
\colhead{$\nu_{\rm max}$} & 
\colhead{$\Delta\nu$} & 
\colhead{Age} & 
\colhead{Modal Age} \\
\colhead{} & 
\colhead{} & 
\colhead{} & 
\colhead{(Deg.)} & 
\colhead{(Deg.)} & 
\colhead{(K)} & 
\colhead{($\log({\rm cm/s}^2)$)} & 
\colhead{(dex)} & 
\colhead{(dex)} & 
\colhead{(dex)} & 
\colhead{(M$_\odot$)} & 
\colhead{(R$_\odot$)} & 
\colhead{($\log({\rm cm/s}^2)$)} & 
\colhead{($\mu$Hz)} & 
\colhead{($\mu$Hz)} & 
\colhead{(Gyr)} & 
\colhead{(Gyr)}
}

\startdata
212498924 & 2M13181597$-$1245237 & 3609527147335455744 & 199.5666 & -12.7566 & 4729 & 1.55 & -1.908 & 0.199 & 0.680 & 0.80 & 13.72 & 2.07 & 14.4 & 2.38 & 11.3 & 8.6 \\
201226802 & 2M12005757$-$0333311 & 3600860727964887680 & 180.2399 & -3.5587 & 4567 & 1.70 & -1.172 & 0.242 & 0.323 & 0.79 & 15.73 & 1.94 & 11.0 & 1.92 & 13.4 & 9.9 \\
228946727 & 2M12390889$-$0227021 & 3682958718591170560 & 189.7870 & -2.4506 & 4849 & 2.09 & -1.814 & 0.218 & 0.186 & 0.81 & 7.17 & 2.64 & 53.1 & 6.35 & 11.9 & 11.0 \\
251545861 & 2M13223901$-$0228562 & 3638105756643622912 & 200.6626 & -2.4823 & 4766 & 2.17 & -1.075 & 0.207 & 0.574 & 0.78 & 10.01 & 2.33 & 26.4 & 3.77 & 14.4 & 13.0 \\
211480777 & 2M08441675$+$1251409 & 602285256783119616 & 131.0698 & 12.8614 & 4598 & 1.81 & -1.038 & 0.159 & 0.409 & 1.09 & 18.97 & 1.92 & 10.4 & 1.71 & 4.4 & 4.2 \\
216439618 & 2M18540793$-$2220424 & 4078727070039971968 & 283.5331 & -22.3451 & 4894 & 2.16 & -1.401 & 0.344 & 0.018 & 0.62 & 9.99 & 2.23 & 20.7 & 3.36 & 31.9 & 29.8 \\
212510240 & 2M13171736$-$1230546 & 3609593083673337472 & 199.3223 & -12.5152 & 4517 & 1.77 & -1.031 & 0.274 & 0.383 & 0.71 & 15.33 & 1.92 & 10.5 & 1.90 & 21.4 & 18.8 \\
203520011 & 2M16110245$-$2551498 & 6043468754447485696 & 242.7602 & -25.8638 & 4706 & 2.11 & -1.144 & 0.315 & 0.500 & 0.68 & 9.12 & 2.35 & 27.7 & 4.04 & 24.5 & 22.0 \\
211326502 & 2M08550079$+$1022483 & 597826737133200256 & 133.7533 & 10.3801 & 5033 & 2.73 & -1.200 & 0.352 & 0.483 & 0.91 & 5.83 & 2.86 & 88.0 & 9.15 & 9.4 & 9.1 \\
... & ... & ... & ... & ... & ... & ... & ... & ... & ... & ... & ... & ... & ... & ... & ... & ... \\
\enddata

\end{splitdeluxetable*}
Table~\ref{tab:k2_poor_table} contains all stars (230) from the APO-K2 catalog with ${\rm [Fe/H]} < -1$. All of these stars are assumed to be RGB stars, as no RC stars should exist in this metallicity range. The masses (and, thereby, ages) of these stars have been calculated using the uncalibrated temperatures from APOGEE, which are those inferred from fitting to a grid of synthetic spectra generated by the spectral synthesis code Turbospectrum \citep{alvarez_plez1998,plez2012} under 1D local thermodynamic equilibrium. These temperatures are shown by \akccite\ to produce masses more in-line with what is expected for halo populations from stellar isochrones, though a mean offset of up to $\approx 10\%$ may remain.

\section{Per-Campaign Age Distributions}
\begin{figure*}
    \includegraphics[width=\textwidth]{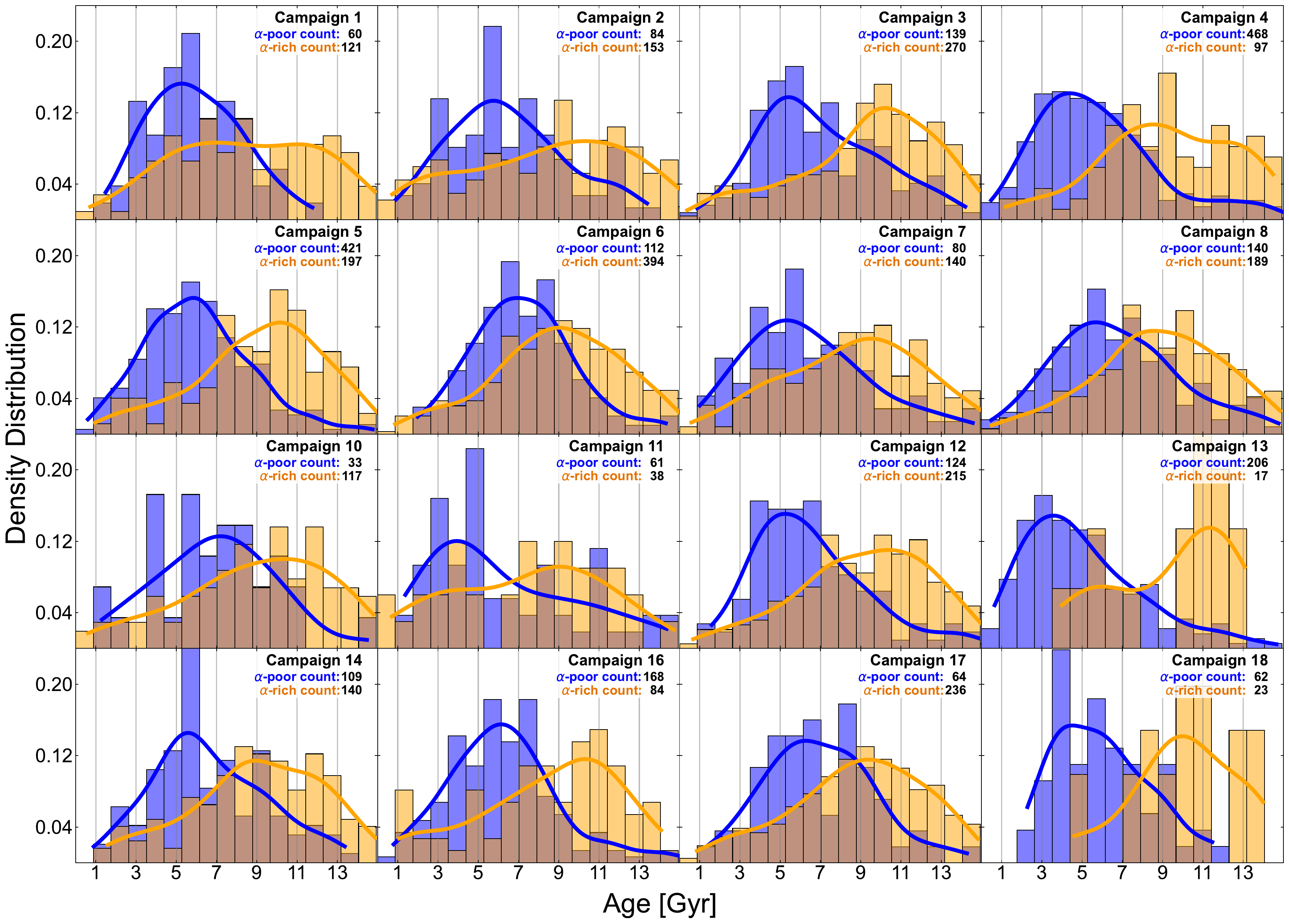}
    \caption{Density distributions for the ages of stars in the $\alpha$-poor (blue) and $\alpha$-rich (orange) RGB populations, split up by K2 campaign. Excluded is Campaign 15, for which we were only able to recover ages for 3 RGB stars. Densities are calculated per chemical population, such that the areas under the $\alpha$-poor and $\alpha$-rich curves both, individually, sum to one, and so the relative heights of these distributions are not reflective of the relative stellar counts in each field. Embedded in the plot, we have also included the number of stars per chemical population per campaign.}
    \label{fig:camps}
\end{figure*}
\begin{figure*}
    \includegraphics[width=\textwidth]{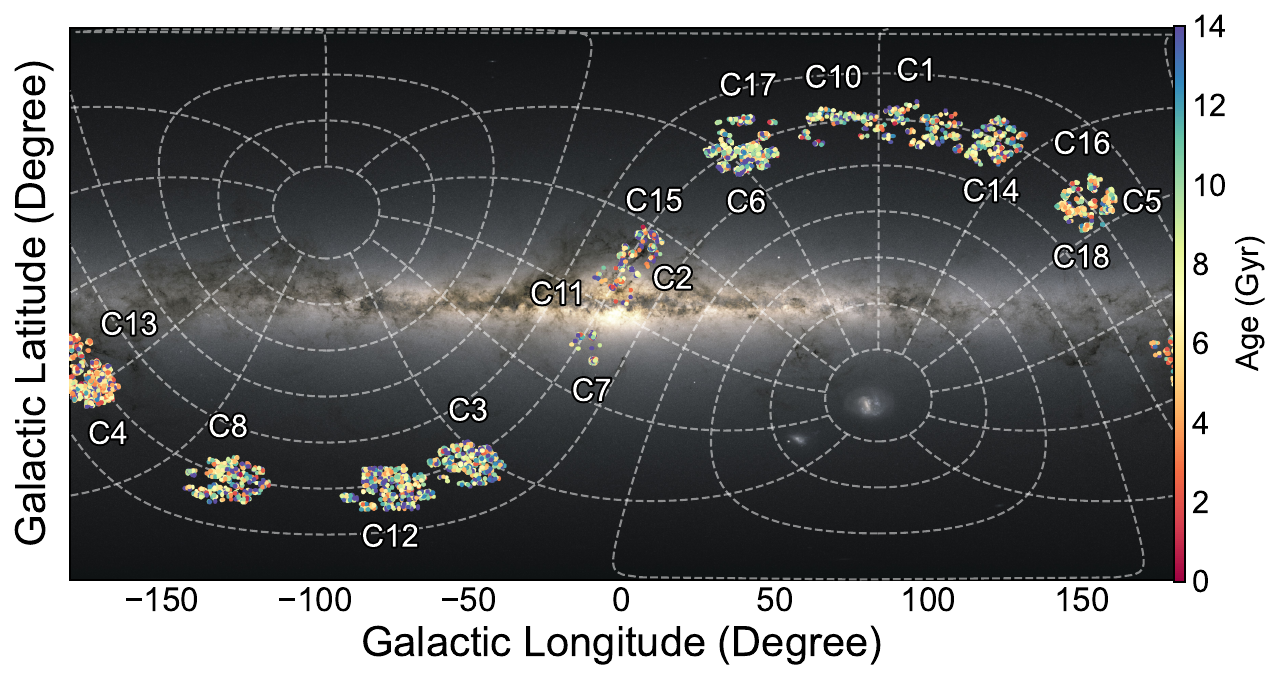}
    \caption{Map of the K2 Campaigns by Galactic Longitude and Latitude. Each point represents an individual star, with each point being colored by age. The background image is modified from ESA/Gaia/DPAC, and is applied using the {\tt mwplot} Python module and the {\tt MWSkyProjection} map "equirectangular".}
    \label{fig:skymap}
\end{figure*}
In Figure~\ref{fig:camps}, we show histograms for the age distributions of the $\alpha$-rich and $\alpha$-poor populations, separated by campaign. In Figure~\ref{fig:skymap}, we map these campaigns on the sky, colored by age. %\jcz when interpreting this I am finding myself wanting a map of where the campaigns are on the sky. I know this is in other K2 papers, but it is very useful and think that plot should be added here, with a reference to it in the main text not just appendix. because it is so clear 


\bibliography{bibliography}{}
\bibliographystyle{aasjournal}



\end{document}